\documentclass[
 amsmath,amssymb,
 aps,
]{revtex4-2}

\usepackage{graphicx}
\usepackage{dcolumn}
\usepackage{bm}
\usepackage{xcolor}

\begin{document}

\preprint{APS/123-QED}

\title{Lagrangian particles and particle clusters in large-aspect-ratio turbulent\\ Rayleigh-B\'enard convection experiment}

\author{Prafulla P. Shevkar}
\email{prafulla-prakash.shevkar@tu-ilmenau.de}
\author{Mitanjali}
\author{Roshan J. Samuel}
\author{Christian Cierpka}%
\author{J\"org Schumacher}
\affiliation{%
 Institute of Thermodynamics and Fluid Mechanics, Technische Universit\"at Ilmenau, Postfach 100565, D-98684 Ilmenau, Germany}%

\date{\today}

\begin{abstract}
We perform Lagrangian particle-tracking experiments in turbulent Rayleigh--Bénard convection at aspect ratio $\Gamma=10$, Rayleigh number $Ra=9.3\times10^7$, and Prandtl number $Pr=4.8$. Approximately 50,000 neutrally buoyant particles are tracked for durations of up to 150 Kolmogorov times. The resulting trajectories are sufficiently long to capture the mean-square displacement of single particles from the short-time ballistic regime to the long-time diffusive regime. We further analyse Lagrangian velocity structure functions and particle-pair dispersion, both of which exhibit scaling behavior consistent with theoretical expectations in their respective temporal regimes. In the Richardson-like regime, the pair-dispersion exponent exceeds 3, most clearly in the lateral direction. Multi-particle statistics are investigated by analysing particle clouds with initial radii of approximately 8 and 20 Kolmogorov lengths using principal component analysis. The clouds spread more strongly in the lateral directions than in the vertical direction and deform from initially spherical shapes into oblate ellipsoids within approximately 10 Kolmogorov times, revealing pronounced anisotropic dispersion. The dynamics is identified from the particle-connectivity network through a spectral analysis of the graph Laplacian followed by $k$-means clustering. Furthermore, we examine the probability density functions of all three acceleration components and assess the statistical convergence of their second and fourth moments. The lateral acceleration components are more intermittent than the vertical component, as indicated by their heavier tails. Acceleration statistics conditioned on the bulk region are slightly less intermittent than those obtained over the full measurement volume, which we attribute to the highly intermittent plume-ejection and plume-aggregation events occurring near the walls.
\end{abstract}

\maketitle

\section{\label{sec:intro}Introduction}
Transport and mixing in turbulent flows has often to be discussed in the Lagrangian frame of reference, which is attached to material lines or surfaces \cite{toschi_2009_lagrangian,Yeung2002Lagrangian}. This is particularly relevant to many geophysical mesoscale processes in the atmosphere and the oceans \cite{atkinson}. Prominent examples are the atmospheric transport of volcanic ash or tephra \cite{Pardini2024} or the accumulation of microplastics in mesoscale eddies in the upper ocean \cite{DiBenedetto2026}. Several fundamental aspects of Lagrangian transport can be suitably studied in large-aspect-ratio thermal convection cells under controlled laboratory conditions -- a paradigm configuration to geophysical turbulence~\cite{Weiss_2024,Moller_Kaufer_Pandey_Schumacher_Cierpka_2022,Shevkar_Samuel_Cierpka_Schumacher_2025}. Analysis is performed where horizontal motion is not constrained by sidewalls. In numerical simulation this is implemented by periodoc boundary conditions whereas in experiments the central domain is observed. Unlike constrained flows in cells with aspect ratio one, the turbulent bulk of high-aspect-ratio cells is characterized by extended large-scale patterns of circulation rolls, which are termed {\em turbulent superstructures}, that gradually change their composition and orientation~\cite{Pandey_superstructures,StevensPRF18, Moller_Kaufer_Pandey_Schumacher_Cierpka_2022}. Turbulence in Rayleigh-B\'enard convection (RBC) is inhomogeneous and anisotropic. The near-wall regions are characterised by the ejection and aggregation of thermal plumes \cite{shevkar_2022_separating_plumes}, followed by successive plume aggregation with increasing distance from the walls~\cite{Shevkar_etal_2025_hierarchial_network}, resulting in enhanced intermittency in these regions~\cite{Shevkar_Samuel_Cierpka_Schumacher_2025}. To capture the complete flow dynamics, it is important to study turbulent convection in high-aspect-ratio cells using full-depth measurements, which will be done in the present analysis.

The primary quantities of interest in Lagrangian turbulence are Lagrangian particle dispersion, Lagrangian structure functions and particle acceleration which we will define later in the text~\cite{Sreeni_Joerg_LagnViews,Biferale2008_LVSF,Gasteuil2007,Ni_Huang_Xia_2012,Shnapp_Liberzon_2018,Varghese2018,Wang_HIT2026,Li2026LagrangianDroplets,toschi_2009_lagrangian}. Although simulations provided access to these quantities~\cite{EmranSchumacher2010_TracerDy}, previous measurements of acceleration and Lagrangian particle dispersion in unity-aspect-ratio cell experiments were limited to the bulk of the cell, or to distances of at least approximately ten thermal boundary-layer thicknesses from the walls and relied on relatively short particle tracks~\cite{Ni_Huang_Xia_2012,Ni_dispersion,Li_Huang_Ni_Xia_2021}. Moreover, Lagrangian particle-dispersion measurements in high-aspect-ratio cells employed longer particle tracks, but were so far performed at relatively moderate Rayleigh numbers~\cite{Weiss_2024}. Furthermore, from the perspective of pollutant transport, the transport dynamics of a cloud of particles needs more attention in laboratory experiments and requires multi-particle statistics beyond one, two or four tracers~\cite{toschi_2009_lagrangian,Schumacher2009}. These are motivating points for the present controlled laboratory experiments of Lagrangian turbulent convection.

In this paper, we investigate the three-dimensional Lagrangian particle statistics in a RBC cell filled with water with a large aspect ratio of $\Gamma=\textrm{width}/\textrm{height}=10$ at a relatively high Rayleigh number of $Ra = g\alpha \Delta T H^3/\nu\kappa=9.27\times10^7$ and Prandtl number of $Pr=\nu/\kappa=4.84$, where $g$ is the acceleration due to gravity, $\alpha$ is the volumetric thermal expansion coefficient, $\Delta T$ is the imposed temperature difference across the cell height $H$, $\nu$ is the kinematic viscosity, and $\kappa$ is the thermal diffusivity.

Particle-tracking measurements are performed over the full depth of the cell in its central region, within a volume of $140\times130\times70$ mm$^3$ located at a normalised distance of at least $x/H=3$ from the sidewalls, using the Shake-the-Box (STB) algorithm~\cite{schanz_2016_shakethebox}, as implemented in DaVis 10.2. Tracer particles are tracked for durations of up to 150 Kolmogorov time. These track lengths exceed the characteristic acceleration- and velocity-correlation time scales and provide a sufficiently long temporal window to resolve tracer displacement from the ballistic to the long-time diffusive regime. They allow us to investigate the short-, intermediate-, and long-time regimes of Lagrangian particle dispersion, including multi-particle dispersion, Lagrangian velocity structure functions, and acceleration statistics in all three spatial directions. The analysis is done in the bulk of the convection layer as well as across the whole height, which will allow us to quantify the impact of the boundary layer dynamics on the overall statistics. The PDFs of the acceleration components obtained from the full-depth measurements are more intermittent than those obtained in the bulk region. The dispersion of particle clouds is found to be more pronounced in the horizontal directions than in the vertical direction, while the extent of the cloud spread remains within the characteristic half-width of the superstructure roll.

Details of the experimental setup and measurements are provided in Section~\ref{sec:expts}. Results on acceleration statistics and Lagrangian particle statistics are presented in Sections~\ref{sec:accln} and~\ref{sec:lagn_Particle_Disp}, respectively. The latter section discusses Lagrangian particle dispersion and velocity structure functions. Single-particle and particle-pair statistics are presented in Sections~\ref{sec:SPS} and \ref{sec:PPS}, respectively. Furthermore, multi-particle statistics based on principal component analysis (PCA) of particle clouds are presented in Section~\ref{sec:multi-particle_stats}. We conclude and provide a brief outlook in Sec. IV. More details of the experimental calibration setup and the track-filtering procedure are provided in appendices.

\section{\label{sec:expts}Particle Tracking Velocimetry}
\begin{figure*}
\centering
{\includegraphics[width=0.8\linewidth,trim=0 0 0 0,clip]{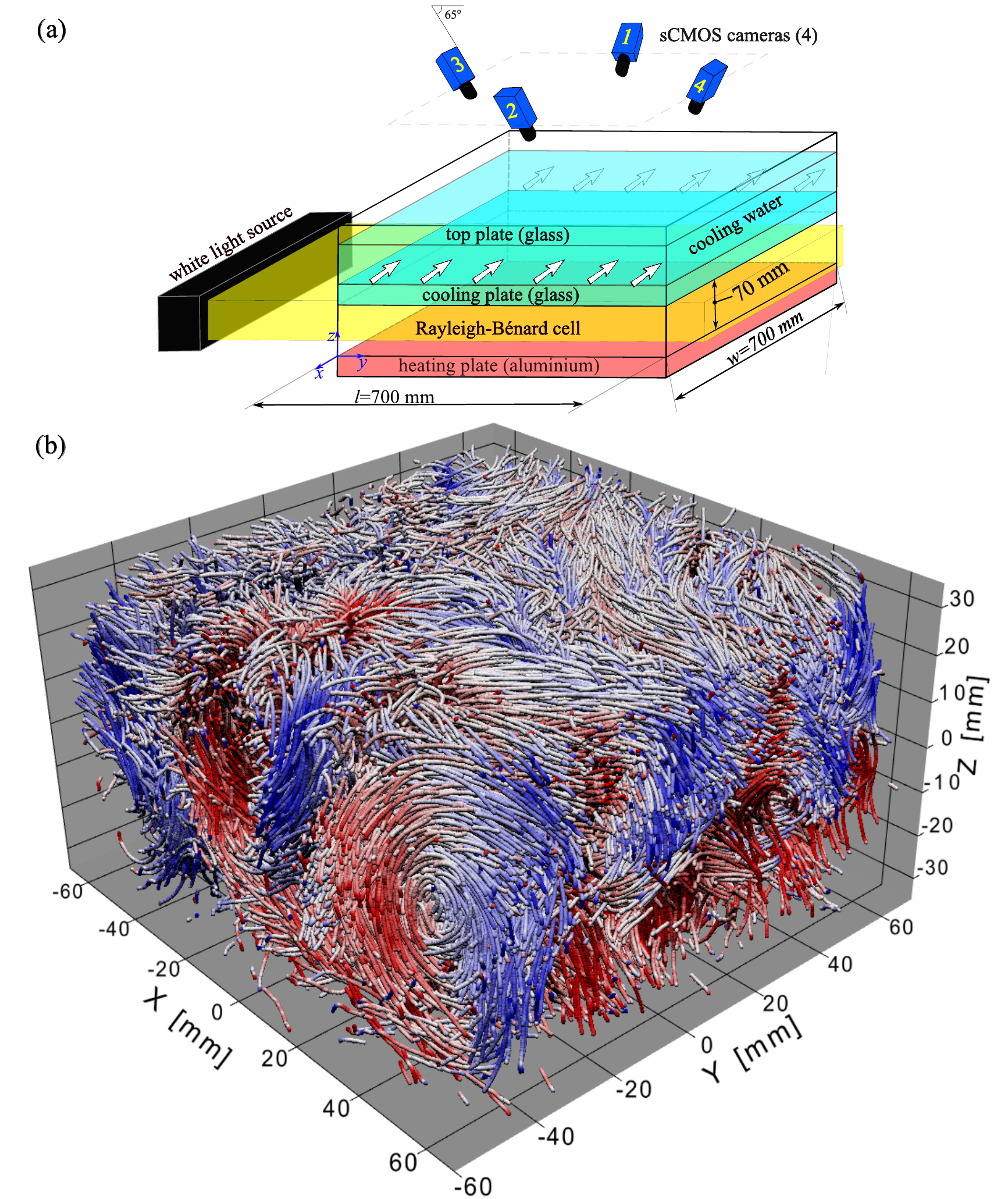}}
\caption{The Rayleigh-B\'enard convection experiment. (a) Sketch of the experimental configuration. (b) Snapshot of particle trajectories in the central fraction of the cell, tracked continuously for at least 100 time instants in the measurement volume at $Ra=9.27\times10^7$ and $Pr=4.84$. Particles are coloured as per their vertical velocity: red and blue denote positive and negative values, respectively, while white denotes near-zero velocity. Different flow patterns, such as upwelling, downwelling, and impingement near the top plate, are clearly visible.}
\label{fig:schematic}
\end{figure*}

Particle tracking experiments were performed to study Lagrangian statistics in a Rayleigh--B\'enard convection cell of $\Gamma=10$. To obtain longer trajectories, measurements were conducted across the full depth of the cell (height $H=70$ mm) and subsequently processed using the STB algorithm~\citep{schanz_2016_shakethebox} available in DaVis~10.2. Further, the raw particle positions obtained from STB were regularised using a penalised-spline approach.

\subsection{\label{sec:setup}Experimental setup}
The experiments were carried out in a flat convection cell with a quadratic cross section at an aspect-ratio $\Gamma=10$ with dimensions $700 \times 700 \times 70$~mm$^3$. A schematic of the experimental setup is shown in Fig.~\ref{fig:schematic}. The cell was filled with deionized, degassed water. The temperatures of the bottom heating plate and top cooling plate were maintained constant to impose a constant temperature difference of $\Delta T = T_h - T_c = 8.58$~K, where $T_h$ and $T_c$ are the mean temperatures of the heating and cooling plates, respectively. Experimental parameters are listed in Table~\ref{tab:expt_para}. We conducted a single run at $Ra=9.27\times10^7$, which is the highest Rayleigh number attainable in the present geometry. This Rayleigh number is nearly an order of magnitude higher than the recent experiments by Weiss et al. in a cell at $\Gamma=8$ ~\cite{Weiss_2024} as well as our previous studies~\cite{Moller_Kaufer_Pandey_Schumacher_Cierpka_2022,Kaeufer2023}. The experimental configuration is similar to that of Shevkar et al.~\cite{Shevkar_Samuel_Cierpka_Schumacher_2025} except that the aspect ratio is decreased to increase $Ra$ by a factor of 20. 

The sidewalls were adequately insulated to minimize heat loss. The temperatures of the heating and cooling plates were maintained by supplying constant-temperature water to their respective circuits from two separate thermostats. The mean temperature of the aluminium heating plate was measured using five embedded PT100 thermistors. The cooling plate is made of glass to allow optical access from the top, and its mean surface temperature was estimated using four thermistors. The maximum absolute deviation among the readings of the individual sensors on the heating and cooling plates was 0.06~K and 1~K, respectively. 
The thermistor mounted near the centre of the heating plate indicated a slightly higher temperature than those located near the sides. Thus, the temperature of the heating plate was locally uniform, with a small temperature gradient from the sides toward the centre existed because of the heating-water distribution network within the heating circuit. Since the actual measurement area accounted for only about 5\% of the total area, the temperature within the measurement region was assumed to remain constant. See also Vieweg et al.~\cite{Vieweg2025} for a detailed discussion of the temperature deviations at both plates. Physical properties of water were estimated at bulk fluid temperature $T_b=(T_h+T_c)/2=34.92~^o$C. Then, the values of control parameters $Ra$ and $Pr$ were found to be $9.27\times10^7$ and $4.84$, respectively. At these values of $Ra$ and $Pr$, calculated Kolmogorov length $\eta_k$ and time $\tau_k$ were 0.66 mm and 0.61 second, respectively~\cite{GROSSMANN_LOHSE_2000}.

\begin{table}[htbp]
\caption{\label{tab:expt_para}%
List of essential experimental parameters. These are the temperatures at the bottom and top, $T_h$ and $T_c$, the temperature difference $\Delta T$, the frames per second (fps), the Rayleigh number $Ra$, the Prandtl number $Pr$, the Kolmogorov length $\eta_k$, the lateral (L) and vertical (V) root mean square (rms) velocities, $u^L_{\rm rms}$ and $u^V_{\rm rms}$, the Kolmogorov time $\tau_k$, free-fall time ($t_f$) and the number of particles $N_p$. The corresponding physical units are given in the head of the table.}
\begin{ruledtabular}
\begin{tabular}{cccccccccccc}
\textrm{$T_h$}&
\textrm{$T_c$}&
\multicolumn{1}{c}\textrm{$\Delta T$}&
\textrm{fps}&
\textrm{$Ra$}&
\textrm{$Pr$}&
$\eta_k$&$u^L_{\rm rms}$&$u^V_{\rm rms}$&
$\tau_k$&$t_f$&
$N_p$\\
$^o$C&$^o$C&$^o$C&Hz&&&\textrm{mm}&&&\textrm{s}&\textrm{s}\\
\colrule
39.20 & 30.62 & 8.58 & 28, 30& $9.27\times10^7$&4.84&0.66&0.075&0.071&0.61&1.56&50000\\
\end{tabular}
\end{ruledtabular}
\end{table}

The flow was seeded with neutrally buoyant polyamide particles of 50~\textmu m in diameter and density $\rho_p = 1.01$~g~cm$^{-3}$. The particles followed the flow faithfully, as the estimated Stokes number was of the order of $10^{-4}$. Tracer particles throughout the depth were illuminated using a pulsed white LED light source, and the scattered-particle images were recorded by four sCMOS cameras (2560$\times$2160 pixels) positioned above the cell. The cameras were equipped with 100~mm focal length lenses (Zeiss Milvus 2/100) via Scheimpflug adapters and mounted symmetrically at an angle of 25$^\circ$ to the vertical axis, as shown in Fig.~\ref{fig:schematic}. The flow was allowed to attain a statistically steady state before measurements were carried out. Particles settled at the bottom plate were redistributed using a 3 mm diameter steel rod, which was moved along cell's length by magnets operating from outside the cell. After steady state was re-established, images were recorded at 28 Hz and 30 Hz, with ten sets in total (one set at 30 Hz), each comprising either 3000 or 4000 frames. The resulting particle image seeding density for recording rate of 28 Hz was about 0.02 particles per pixel while that for recording rate of 30 Hz was 0.008.  

The calibration was carried out using 3D calibration plate (No, 204-15, LaVision GmbH) by providing two views separated by a distance of 36.8 mm. A polynomial function was used for the fit, and the maximum fitting error among both views for any camera was less than 0.4~pixels. The details of calibration setup and procedure are given in Appendix~\ref{sec:calibration}.

\subsection{\label{sec:measurements}Volume self-calibration and Shake-The-Box processing}
The recorded images were first pre-processed, and a geometric mask was applied to extract the common field of view shared by all four cameras. The resulting cropped images were, subsequently, used for volume self-calibration \citep{wieneke2008}. To this end, the measurement volume was subdivided into smaller sub-volumes, and the calibration was iteratively refined. The procedure was initiated with sub-volumes of size $3 \times 3 \times 2$ and progressively continued up to a final subdivision of $9 \times 9 \times 7$. For some sets final subdivision of $9 \times 9 \times 5$ was used. In the final pass, an allowable triangulation error was 1 voxel. At the end of the volume-self-calibration, the average disparity across the entire measurement volume was reduced down to below 0.1 voxel. Finally, the optical transfer function (OTF) of the particle imaging system was calibrated \citep{Schanz2012_OTF}. Several subdivisions in the $z-$direction also helped to account for the changing particle image size and intensities due to oblique imaging and difference in depth levels.

The resulting calibration was then used to process the time-resolved particle images using the STB algorithm to determine the 3D particle positions in the measurement volume for the dataset. A variable time-step processing approach was employed, as it is known to minimize ghost tracks and improve tracking accuracy~\cite{Schanz2021_VTSTB}. The velocity limits in the $x$- and $y$-directions were set according to the maximum particle displacement per unit time in the horizontal plane. The limit in the $z$-direction was determined by trial and error, starting from the value used for the $x$- and $y$-directions. The acceleration limits were likewise determined by trial and error, beginning with an initial value of 1 voxel~\citep{Shevkar_Samuel_Cierpka_Schumacher_2025}. During the final processing, only tracks with lengths greater than or equal to 10 time steps were accepted, resulting in about $N_p\approx 50,000$ tracked particles at any given instant within the measurement volume.

\subsection{Post-processing of Lagrangian tracer tracks}
The raw particle positions obtained from Shake-The-Box post-processing are contaminated by noise from multiple sources and therefore require post-processing. Only tracks longer than 50 time steps are considered for post-processing and used in the analysis. The Lagrangian tracks were regularised using a P-spline (Penalised-spline) approach~\cite{Eilers1996PSplines} implemented in MATLAB, similar to the TrackFit method~\cite{Gesemann2016NoisyParticleTracks}. Cubic B-splines were fitted separately to the particle positions in each direction, with a penalty imposed on their third derivatives (jerk). The regularised data were subsequently evaluated at the original data points. The regularisation parameters were determined from the cut-off frequencies obtained from the power spectral density (PSD) analysis of the position signals. We use the cut-off frequency corresponding to a signal-to-noise ratio of unity. This cut-off yields acceleration errors close to the minimum mean acceleration error~\cite{Cheminet2021RegularizedBSplineSmoothing}. Noise characteristics differ between the lateral and vertical directions, as well as between short and long tracks. The latter have a length of greater or equal than 500 time-steps, thus separate cut-off frequencies were used for these categories. More details are given in Appendix section~\ref{sec:psd}. 

\subsection{\label{sec:autocorrelationFn}Velocity and acceleration autocorrelation functions}
\begin{figure*}
\centering
{\includegraphics[width=0.48\linewidth,trim=0 0 0 0,clip]{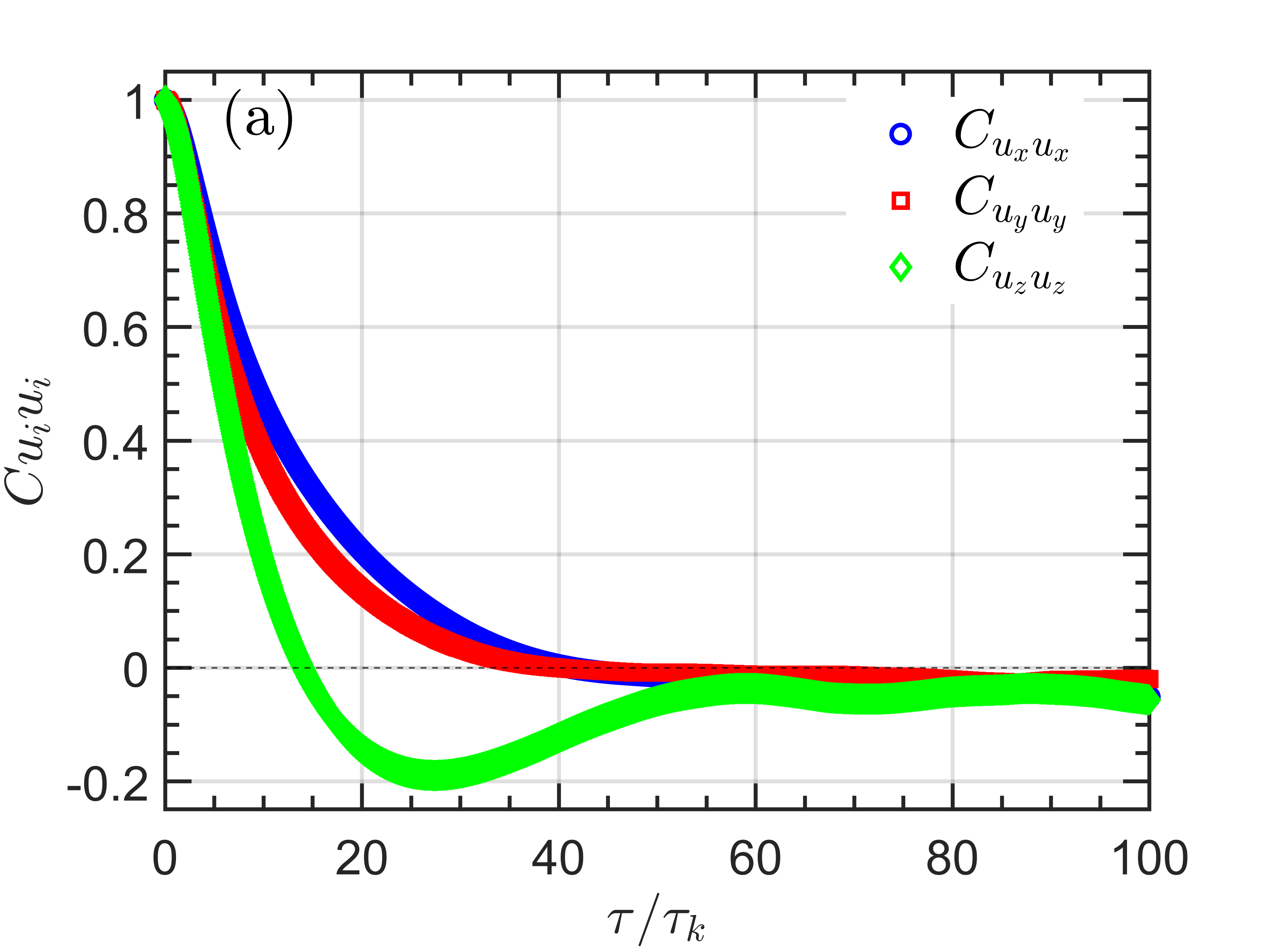}}
{\includegraphics[width=0.48\linewidth,trim=0 0 0 0,clip]{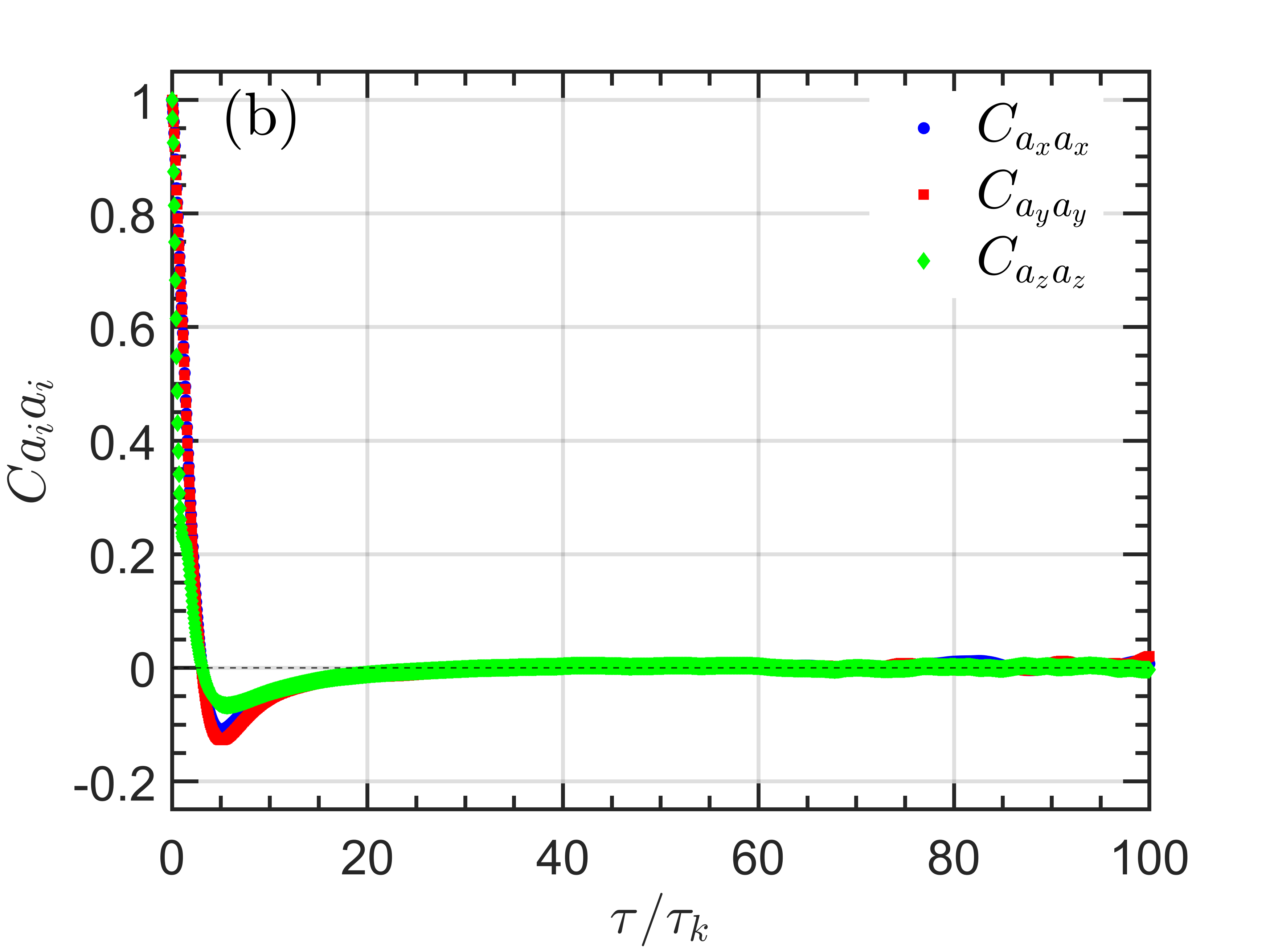}}
\caption{Normalised velocity autocorrelation functions for $u_x$, $u_y$, and $u_z$ in (a), and normalised acceleration autocorrelation functions for $a_x$, $a_y$, and $a_z$ in (b).}
\label{fig:autocorr}
\end{figure*}
We now compute the velocity and acceleration autocorrelation functions and show that the particle-track lengths are sufficiently long to capture the dynamics associated with the relevant turbulent time scales. In general, an autocorrelation function measures how strongly a quantity at a given time remains correlated with its value at an earlier time. The normalised autocorrelation function of the velocity component for a time lag $\tau$ is defined as
\begin{equation}
    \label{eq:Cuu}
    C_{u_i u_i}(\tau)
    =
    \frac{
    \left\langle u_{i}(t)u_{i}(t+\tau)\right\rangle_{t,p}
    }{
    \left\langle u_{i}^2(t)\right\rangle_{t}
    }
    \qquad\mbox{with}\quad i=\{x,y,z\}\,,
\end{equation}
where $u_i(t)$ is the $i$-component of the particle velocity at time $t$, and $\langle\cdot\rangle_{t,p}$ denotes averaging over time and over all particles. Figure~\ref{fig:autocorr}(a) shows the Lagrangian velocity autocorrelation functions for all three velocity components, plotted against normalised time. The autocorrelation function of the vertical velocity component rapidly decays to zero, becomes negative, and then gradually returns to zero at approximately $\tau/\tau_k = 60$. This behaviour indicates that particles trapped within a superstructure pattern lose their vertical-velocity correlation as they approach the plate. The correlation briefly becomes negative as the particles reverse direction and continue along their trajectories within the superstructure, before gradually decaying to zero while exhibiting weak oscillations about zero. In contrast, the autocorrelation functions of the horizontal velocity components decay more slowly and subsequently show weak oscillations around zero. The eventual loss of correlation indicates that the particle trajectories are sufficiently long to capture the relevant dispersion regimes. Moreover, all autocorrelation functions approach zero by $\tau/\tau_k \approx 60$ and show only small oscillations thereafter, indicating that the flow has become effectively uncorrelated with its initial state.

The normalised autocorrelation function of the acceleration component for a time lag $\tau$ is defined as
\begin{equation}
    \label{eq:CAA}
    C_{a_i a_i}(\tau)
    =
    \frac{
    \left\langle a_i(t)a_i(t+\tau)\right\rangle_{t,p}
    }{
    \left\langle a_i^2(t)\right\rangle_{t,p}
    }
    \qquad\mbox{with}\quad i=\{x,y,z\}\,,
\end{equation}
where $a_i(t)$ is the $i$-component of the particle acceleration at time $t$, and $\langle\cdot\rangle_{t,p}$ again denotes an average over time and all particles. Figure~\ref{fig:autocorr}(b) shows the normalised acceleration autocorrelation functions for all three components. The acceleration autocorrelations decay on substantially shorter timescales than the velocity autocorrelations. The autocorrelation of the acceleration components crosses zero at approximately $\tau/\tau_k \approx 3.4$. Importantly, this rapid decay is resolved by approximately 40--50 recorded data points in our experiments. At longer time lags, the autocorrelations gradually approaches zero, reaching negligible values for $\tau/\tau_k > 20$. The autocorrelation functions of the two horizontal acceleration components coincide and reach larger negative values than the one of the vertical component. The difference indicates stronger short-time acceleration reversals or oscillations in the horizontal directions than in the vertical direction. This behaviour may be attributed to the unconstrained horizontal flow in our experimental cell. In contrast, in the unity-aspect-ratio experiments, the autocorrelation functions of the $x$, $y$, and $z$ acceleration components were found to overlap~\cite{Ni_Huang_Xia_2012}.

\section{\label{sec:results}Results}
In the following, we will discuss the results of our experiment. Two independent experimental runs were conducted under identical conditions. The cumulative recording duration was approximately 660 free-fall times $T_f=\sqrt{H/g\alpha\Delta T}$, thus providing sufficient data for reliable statistical analysis. We first present the acceleration statistics in subsection~\ref{sec:accln} and subsequently discuss Lagrangian particle statistics, including single-particle and  particle-pair statistics in subsection~\ref{sec:lagn_Particle_Disp} and multi-particle statistics plus cluster analysis in subsection~\ref{sec:multi-particle_stats}.

\subsection{\label{sec:accln}Acceleration statistics}
\begin{figure*}[htbp]
\centering
{\includegraphics[width=0.85\linewidth,trim=0 0 0 0,clip]{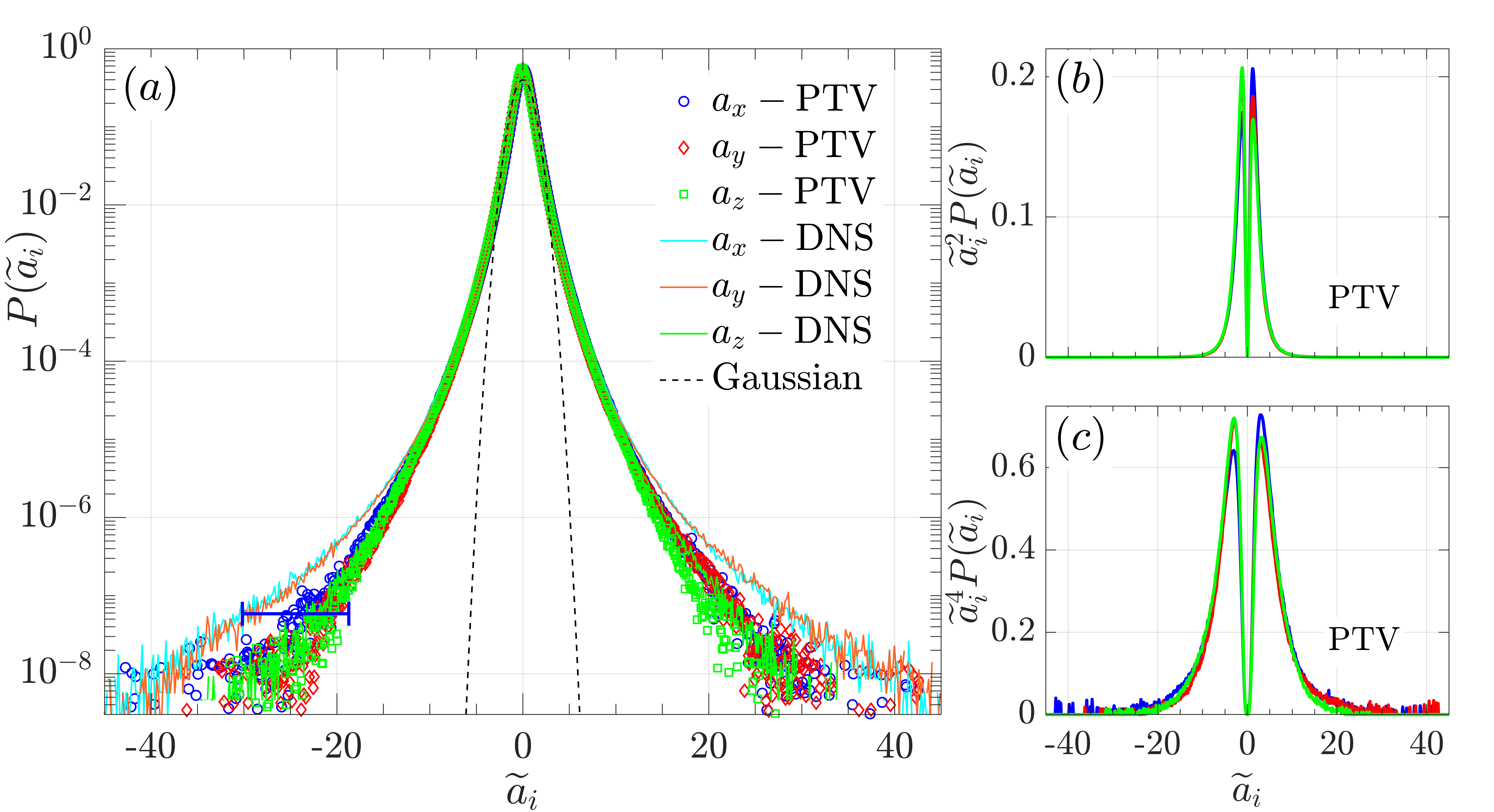}}
\caption{Probability density functions (PDF) of Lagrangian acceleration components. (a) PDFs of the normalized acceleration components $\widetilde{a}_x$, $\widetilde{a}_y$, and $\widetilde{a}_z$. The contributions to the second and fourth moments of the acceleration components are shown in panels (b) and (c), respectively. They underline a converged statistics up to moment order $n=4$. A total of $7.3\times10^8$ events were used to generate the plots. The experimental particle tracking velocimetry (PTV) data are compared with direct numerical simulations (DNS) at exactly the same $Ra$, $Pr$ and $\Gamma$.}
\label{fig:accln_PDFs}
\end{figure*}

Figure~\ref{fig:accln_PDFs} shows the probability density functions (PDFs) of the acceleration components and their higher-order moments. Panel (a) presents the PDFs of the acceleration components normalised by their respective root-mean-square values averaged over volume and time as 
\begin{equation}
    \widetilde{a}_i=\frac{a_i}{\sqrt{\langle a_i^2\rangle}_{V,t}}.
\end{equation}

A total of $7.3\times10^8$ acceleration events were used to construct these PDFs and results are obtained from full-depth measurements. A Gaussian distribution is also shown for comparison. The tails of the measured distributions deviate strongly from the Gaussian distribution, indicating pronounced heavy-tailed behaviour. The PDFs of all acceleration components show this stretched-exponential behaviour. The PDF of the vertical acceleration component, $a_z$, obtained from PTV agrees well with the corresponding direct numerical simulations (DNS) result, including in the far tails. The PDFs of the lateral acceleration components display a fair agreement with DNS data within the measurement uncertainty of the PTV data. The maximum possible uncertainty in the lateral acceleration measurement, estimated from the standard deviation of the positional uncertainty using central finite-difference error propagation, is represented by an error bar plotted at one of the data points in the tail. This error bar reaches to the tail of the numerical data. In both investigations, PTV and DNS, the PDFs of the lateral acceleration components are more intermittent than that of the vertical component, consistent with the DNS in Schumacher~\cite{SchumacherPRL2008}. In PTV, the lateral components extend to normalised acceleration magnitudes of approximately 40, where the probability density function decreases to values as low as $10^{-8}$.

Figures~\ref{fig:accln_PDFs}(b) and \ref{fig:accln_PDFs}(c) show the contributions to the second- and fourth-order moments of the components of $\widetilde{a}_i$, respectively. To this end, we plot $\widetilde{a}_i^n P(\widetilde{a}_i)$ versus $\widetilde{a}_i$ for $n=2,\,4$. The area content below these curves corresponds to the normalized moment of corresponding order. These results indicate that the PDF tails are sufficiently well converged to allow reliable estimation of moments up to fourth order. The vertical acceleration component is slightly negatively skewed, whereas the two horizontal acceleration components exhibit skewness of opposite signs. We compute the skewness and flatness of all three components of the normalised acceleration, and the corresponding values are listed in Table~\ref{tab:skew_flat}. The vertical acceleration component shows negative skewness, consistent with the findings in ref.~\cite{SchumacherPRL2008}. The skewness of $a_x$ is slightly positive, while that of $a_y$ is nearly zero. The flatness values of all three components are approximately 9, substantially exceeding the Gaussian reference value of 3. This indicates pronounced heavy-tailed intermittent behaviour.

To further highlight the differences between the bulk and full-depth measurements, we compare the full-depth results with those obtained from the central one-third-height region, which is symmetric about the mid-height of the cell. We also compute the skewness and flatness of all three components of the normalised acceleration in the bulk region, and values are listed in Table~\ref{tab:skew_flat}. The PDFs measured in the bulk are weakly skewed, similar to the nearly symmetric distributions reported by \cite{Ni_Huang_Xia_2012}. The skewness values of $a_x$, $a_y$, and $a_z$ are comparable to the corresponding values obtained from the full-depth measurements. In general, the flatness values obtained over the full depth are larger than those in the bulk region. These higher flatness values are attributed to the relatively stronger fluctuations associated with near-wall plume-ejection and plume-aggregation events~\cite{Shevkar_etal_2025_hierarchial_network}. They cannot be resolved in the present laboratory experiments. These unresolved far tails may have caused from the approximately $5-10\%$ underestimation of the root-mean-square values in the experiments relative to the DNS. Accurately capturing the far tails of the distributions remains a challenge in Lagrangian particle tracking measurements~\cite{Sciacchitano2026SecondLPTChallenge}.

To conclude this subsection, the acceleration statistics, which is very sensitive to intermittency effects, could be resolved with sufficient statistical convergence and is found to agree well and fairly well in the vertical and lateral directions, respectively. Height-dependent analysis reveals an enhanced intermittency in the near-wall region.

\begin{table}[htbp]
\caption{\label{tab:skew_flat}%
Comparison of the skewness and flatness values of the acceleration-component PDFs obtained from the full-depth measurements and the bulk one-third region.}
\begin{ruledtabular}
\begin{tabular}{ccccccc}
Region& \multicolumn{3}{c}{Full-depth}
& \multicolumn{3}{c}{Bulk One-third Region} \\
& \multicolumn{3}{c}{$\textrm{-35 mm} < z < 35 \textrm{mm}$}
& \multicolumn{3}{c}{$\textrm{-12 mm} < z < 12 \textrm{mm}$}\\
& $a_x$ & $a_y$ & $a_z$
& $a_x$ & $a_y$ & $a_z$ \\
\colrule
\colrule
\textrm{Skewness} & 0.133 & -0.051 &-0.105& 0.151&-0.055&-0.108\\
\textrm{Flatness} & 9.65 & 9.01 &9.18& 8.71&8.31&9.31\\
\end{tabular}
\end{ruledtabular}
\end{table}

\subsection{\label{sec:lagn_Particle_Disp}Single-particle and particle-pair statistics}
In the current experiment, particles could be tracked for times up to $150\tau_k$. The implemented particle-seeding density employed in the experiments yielded a sufficient number of particle pairs with initial separations of $\eta_k$ or less. These measurements enabled us to investigate single-particle and particle-pair dispersion and statistics across a broad range of temporal scales, from the short-time ballistic regime, through intermediate inertial-range scales, to the long-time diffusive regime~\cite{taylor_1922_diffusion,richardson_1926_atmospheric,batchelor_1950_similarity,batchelor_1952_relative}.

\subsubsection{\label{sec:SPS}Single-particle statistics}

\begin{figure*}
\centering
{\includegraphics[width=0.495\linewidth,trim=0 0 0 0,clip]{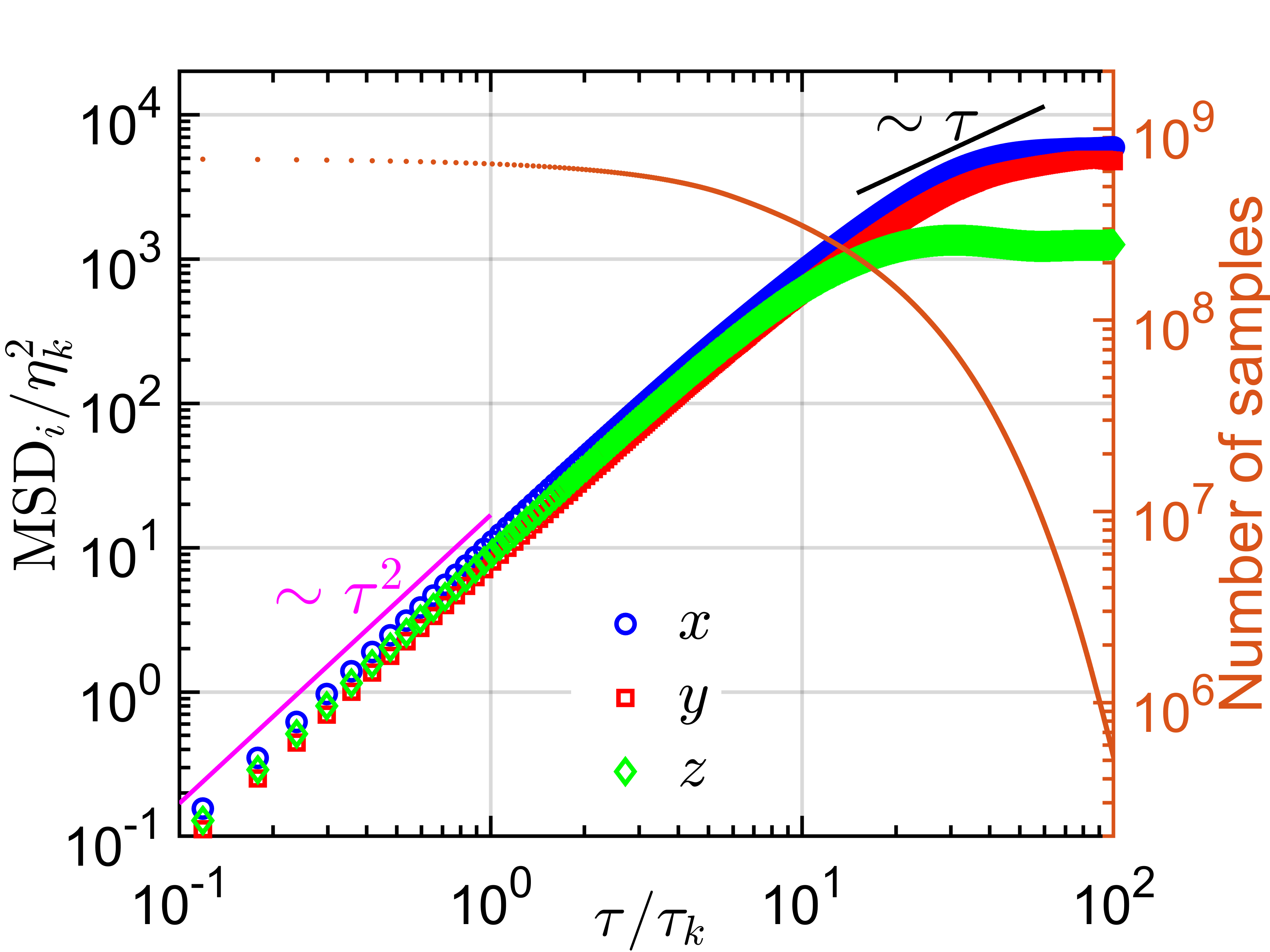}}
{\includegraphics[width=0.495\linewidth,trim=0 0 0 0,clip]{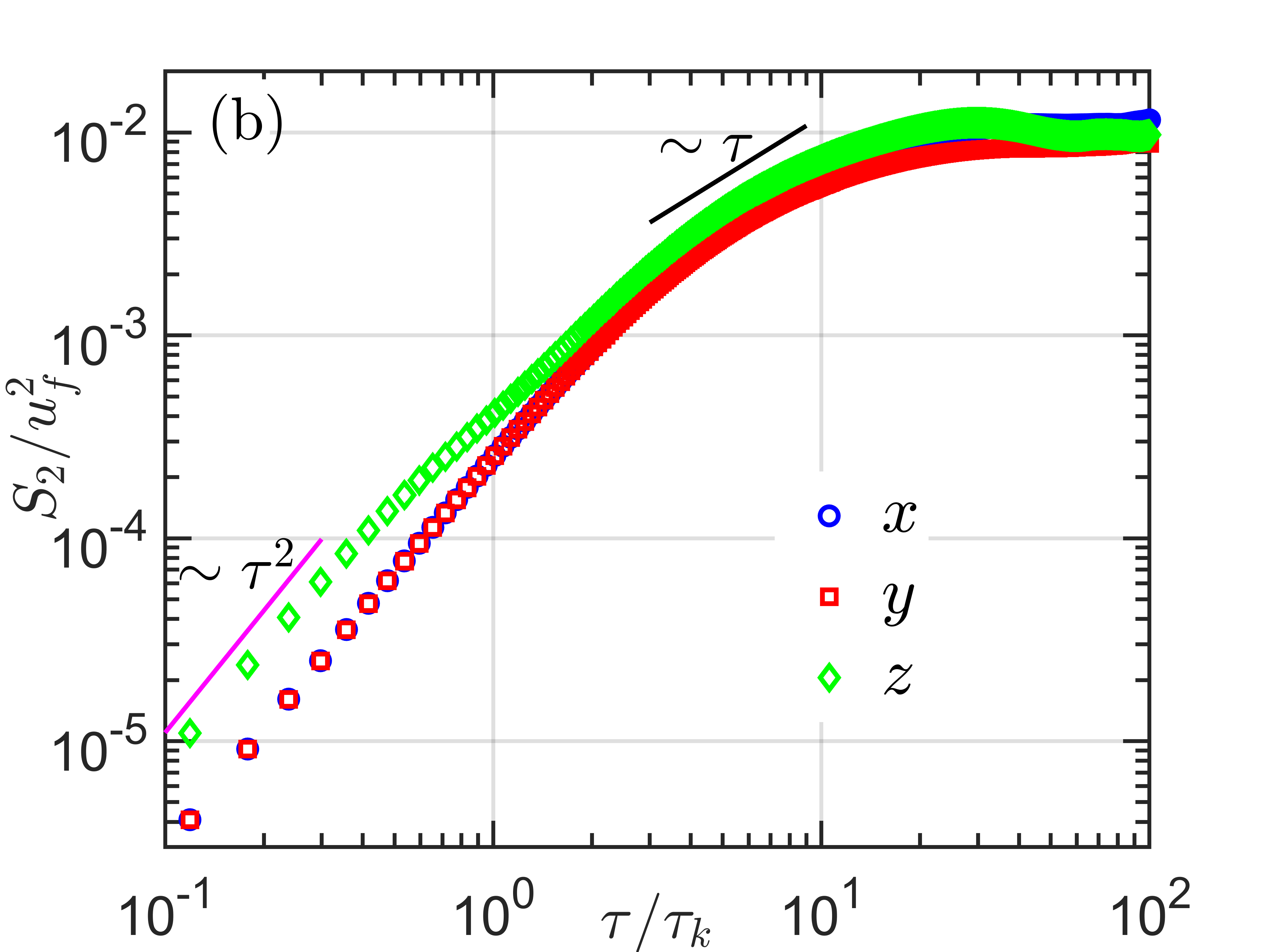}}\\
{\includegraphics[width=\linewidth,trim=0 0 0 0,clip]{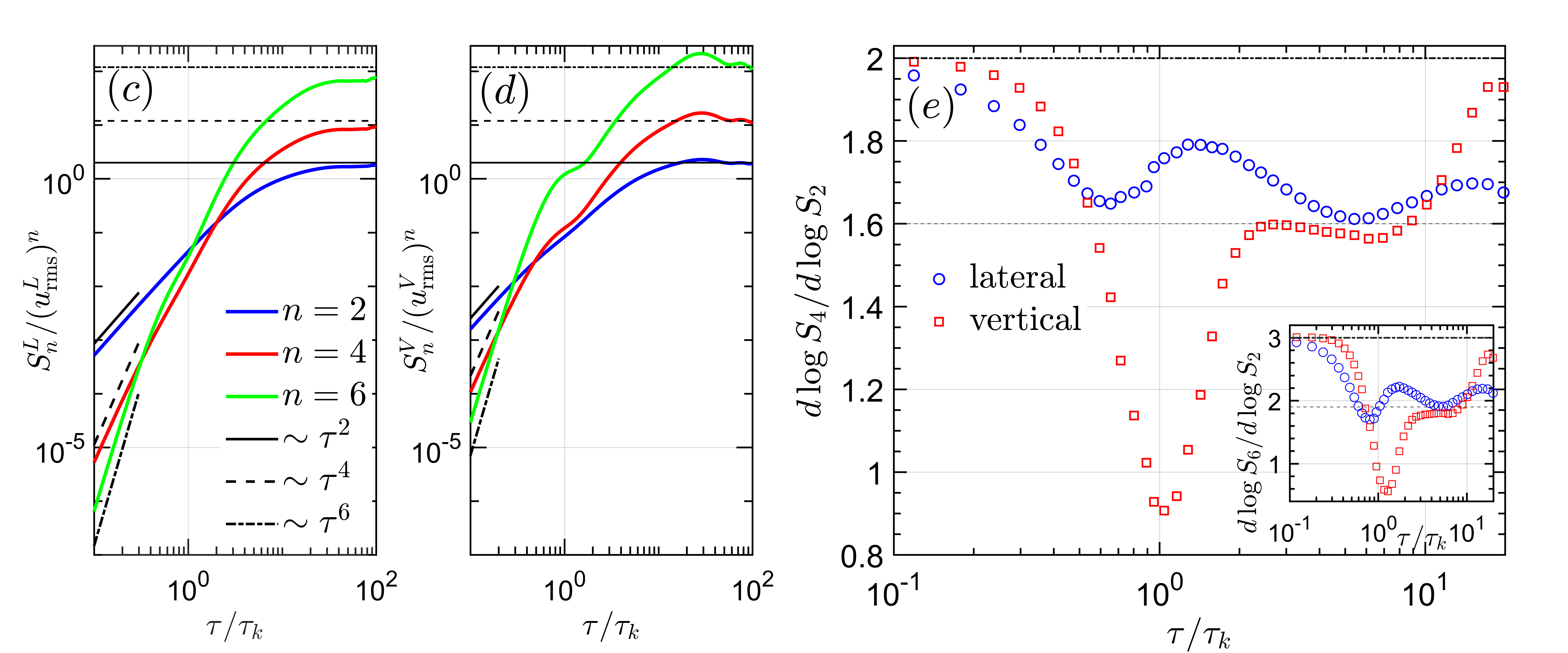}}
\caption{Time evolution of (a) the normalised mean-square particle displacements in the lateral and vertical directions, (b) the normalised second-order Lagrangian velocity structure functions (LVSFs) for the lateral and vertical velocity components, (c) the higher-order LVSFs for the lateral velocity component, (d) the higher-order LVSFs for the vertical velocity component, and (e) the fourth-order local scaling exponent. The Gaussian decorrelation limits for the second-, fourth-, and sixth-order structure functions are shown at values of 2 (solid line), 12 (dashed line), and 120 (dash-dotted line), respectively. The inset of panel (e) shows the sixth-order local scaling exponent. The dash-dotted line in in the main panel (e) and the inset indicate the corresponding non-intermittent reference values for the fourth- and sixth-order local scaling exponents, respectively.}
\label{fig:single_particle_disp}
\end{figure*}

Single-particle dispersion characterizes the displacement of a particle over time within the measurement domain in the lateral and vertical directions. For a given time lag $\tau$, the mean-square displacement is defined as
\begin{equation}
    \label{eq:MSD}
{\rm MSD}_i(\tau)=\left\langle \left[q_i(t+\tau)-q_{i}(t)\right]^2 \right\rangle_{t,p}
\end{equation}
where $q_{i}(t)$ is the position of particle in the $i$-direction at time $t$, and $\langle \cdot \rangle_{t,p}$ denotes averaging over time and over all tracked particles to improve statistical convergence. Figure~\ref{fig:single_particle_disp}(a) shows the mean-square displacement MSD of a single particle as a function of time for each of the three spatial components. At short times ($\tau/\tau_k<1$), while the particle velocity remains strongly correlated, the MSD of the horizontal components exhibits ballistic scaling, i.e., $\mathrm{MSD}_{x,y}\sim\tau^2$. At times longer than the velocity-correlation time, approximately $\tau/\tau_k=60$, the horizontal MSD transitions to diffusive scaling, i.e., $\mathrm{MSD}_{x,y}\sim\tau$, reflecting the loss of memory of the initial velocity. The vertical component also show clear ballistic scaling at short times, with $\mathrm{MSD}_z\sim\tau^2$. However, rather than passing through a distinct diffusive regime, $\mathrm{MSD}_z$ rapidly approaches a plateau, indicating that vertical displacement is more strongly restricted by confinement. The number of samples used at each value of $\tau/\tau_k$ is represented by the dotted line, with the corresponding values shown on the right-hand vertical axis. More than $6\times10^8$ samples were used in the shortest time lag, whereas about $5\times10^5$ samples remained at the highest value of $\tau/\tau_k$.

The Lagrangian velocity structure function (LVSF) describes the statistical moments of the velocity fluctuations experienced by a fluid particle as it moves with the flow over a time lag $\tau$. The $n^{\rm th}$-order Lagrangian velocity structure function is defined as
\begin{equation}
    \label{eq:SF}
    S_n(\tau)=\langle |{\bm u}(t+\tau)-{\bm u}(t)|^n \rangle_{t,p}
\end{equation}
Figure~\ref{fig:single_particle_disp}(b) shows the individual second-order LVSFs for all three velocity components in units of the free-fall velocity, where $u_f=H/T_f$. At short time lags, all three LVSFs exhibit quadratic scaling, $S_2\sim(\tau/\tau_k)^2$. For $\tau/\tau_k>1$, all three components progressively depart from this quadratic regime and subsequently show an approximately linear scaling in the inertial subrange, $S_2\sim\tau$, for a very short range. At larger time lags, all three LVSF approach a plateau with a normalised value of approximately $10^{-2}$. In this regime, velocities separated by the time lag become effectively uncorrelated; therefore, $S_2^2 \rightarrow 2\langle u_i^2\rangle$. The normalised LVSF value of approximately $10^{-2}$ exceeds that reported in Ref.~\cite{Weiss_2024} for $\Gamma=16$ cell at lower $Ra$, indicating a higher fluctuation level at higher $Ra$. 

The LVSFs of the two horizontal velocity components, nearly overlap. At short time lags, the vertical component is slightly larger than the horizontal components. Moreover, its quadratic scaling persists over a shorter range of time lags than that of the horizontal components. These differences gradually diminish, and all three components attain comparable values for $\tau/\tau_k>1$. Within this range, the LVSFs of all three velocity components show the same scaling behaviour. Clearly the scaling range of the LVSFs remains small at obtained Ra. A scaling of these quantities is generally more difficult as discussed for example by Yeung~\cite{Yeung2002Lagrangian}. As expected, the earlier scaling transition of the LVSFs relative to the MSD indicates that velocity increments respond rapidly to small-scale turbulent fluctuations, whereas the particle displacement retains memory of the initial velocity over a longer timescale. This is also consistent with the acceleration autocorrelation decaying to zero over a shorter timescale than the velocity autocorrelation. Since the second-order structure functions for the $x$- and $y$-velocity components overlap closely, we combine them as the lateral component for the subsequent analysis.

The Lagrangian velocity along the particle tracks can be decomposed into 
\begin{equation}
    \label{eq:SF1}
    {\bm u}={\bm u}^L + {\bm u}^V={\bm u}^L+u_z {\bm e}_z\,,
\end{equation}
such that definition \eqref{eq:SF} can be refined to
\begin{equation}
    \label{eq:SF2}
    S^L_n(\tau)=\langle |{\bm u}^L(t+\tau)-{\bm u}^L(t)|^n \rangle_{t,p}
\quad\mbox{and}\quad
    S^V_n(\tau)=\langle |u_z(t+\tau)-u_z(t)|^n \rangle_{t,p}\,.
\end{equation}
We will denote them as the lateral (L) and vertical (V) Lagrangian velocity structure functions. For large time lags the LVSF of order $n$ converge to a Gaussian value, which is given by (see e.g. ref. \cite{Konstandin2012})
\begin{equation}
    \label{eq:SF3}
    \lim_{\tau\to \infty} \frac{S^{L,V}_n(\tau)}{\left(u^{L,V}_{\rm rms}\right)^n}= 2^n\frac{\Gamma\left(\frac{n+1}{2}\right)}{\sqrt{\pi}}=2, 12, 120 \quad\mbox{for orders}\quad n=2,\,4,\,6\,.
\end{equation}
We plot the evolution of the second-, fourth-, and sixth-order LVSF moments for the lateral and vertical velocity components in Figs.~\ref{fig:single_particle_disp}(c) and (d), respectively. The structure functions are normalized by the corresponding root-mean-square velocity raised to the respective order. For small time lags, $\tau/\tau_k<1$, the second-, fourth-, and sixth-order structure functions scale approximately as $\tau^2$, $\tau^4$, and $\tau^6$, respectively, as expected in the differentiable regime. At intermediate time lags, the structure functions deviate from these small-scale power-law scalings, with the higher-order moments exhibiting pronounced curvature, particularly for the vertical component. For the longest time lags, the structure functions approach the Gaussian limits given by eq. \eqref{eq:SF3}. These Gaussian decorrelation limits are indicated by the solid, dashed, and dash-dotted lines in the panels.

The statistics of the velocity increments can be further characterised by examining the fourth- and sixth-order moments relative to the second-order moment. The local scaling exponents of order $n$ are then computed from the derivatives of the lateral or vertical $\log S_n(\tau)$ with respect to lateral or vertical $\log S_2(\tau)$. In detail, the local scaling exponents are defined as
\begin{equation}
\zeta_n^{L,V}(\tau)
=
\frac{\mathrm{d}\log S^{L,V}_n(\tau)}
     {\mathrm{d}\log S^{L,V}_2(\tau)} .
\label{eq:local_exponent}
\end{equation}
For the lateral and vertical velocity components, the local scaling exponents $\zeta_4^{L,V}(\tau)$ and $\zeta_6^{L,V}(\tau)$ are computed for a range of time lags $\tau$ and are shown in Fig.~\ref{fig:single_particle_disp}(e). The trends observed in our experiments are qualitatively consistent with those reported in numerical simulations by Biferale et al.~\cite{Biferale2008_LVSF} and Arn\'{e}odo et al.~\cite{Arneodo2008}. At time lags $\tau/\tau_k \ll 1$, $\zeta_4(\tau)$ and $\zeta_6(\tau)$ take the corresponding non-intermittent values of 2 and 3, respectively. At time lags of order $\tau/\tau_k\sim 1$, both exponents show a pronounced dip. Our flow is anisotropic, which manifests in the differences between lateral and vertical local exponents. Subsequently, the lateral and vertical exponents approach a plateau-like region for both orders; this is observed for $\tau/\tau_k\sim 10$. The plateau values are approximately $\zeta^{L,V}_4(\tau)\approx 1.6$ and $\zeta_6^{L,V}(\tau)\approx 1.9$, respectively, thus indicating intermittent Lagrangian velocity fluctuations. These values are close to the values of $1.6\pm0.1$ and $2.0\pm0.1$, respectively, reported by DNS of Biferale et al.~\cite{Biferale2008_LVSF}. The small scaling plateau is more clearly developed for the vertical component than for the lateral one. At longer time lags, fluctuations in the exponent values are observed, likely due to insufficient statistical sampling.

\subsubsection{\label{sec:PPS}Particle-pair statistics}
\begin{figure*}
\centering
{\includegraphics[width=0.48\linewidth,trim=0 0 0 0,clip]{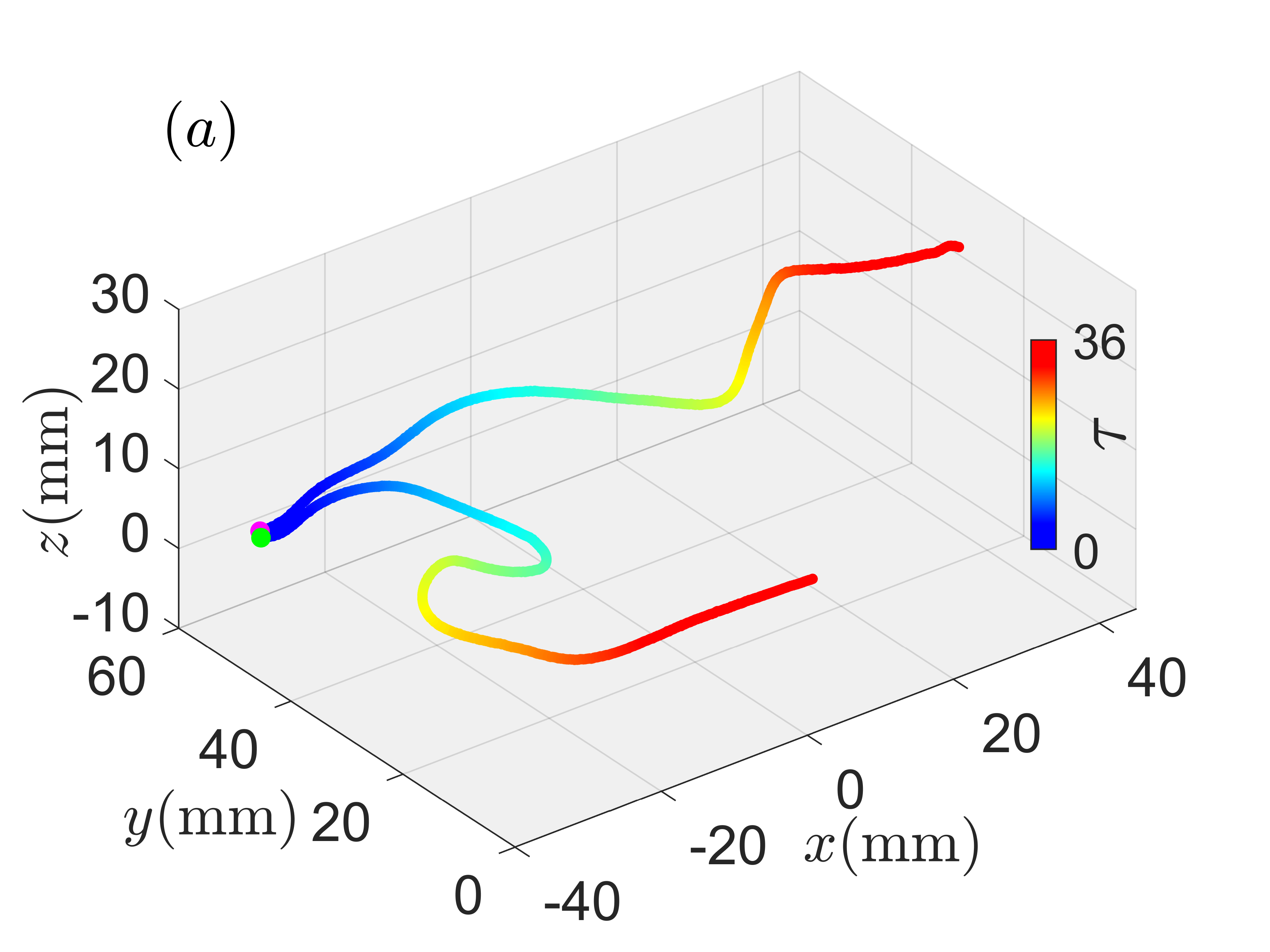}}
{\includegraphics[width=0.48\linewidth,trim=0 0 0 0,clip]{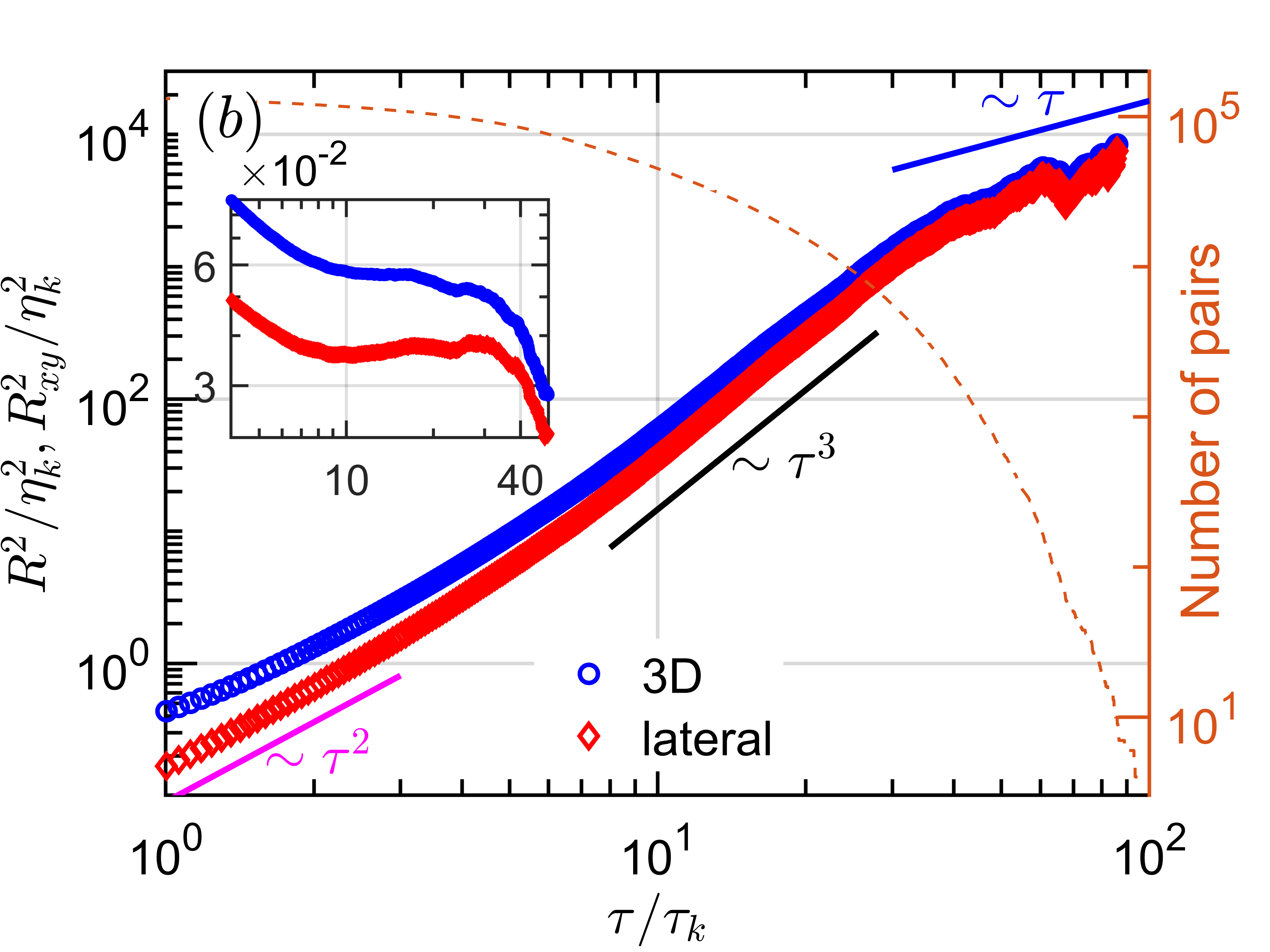}}
\caption{(a) Typical example of a particle pair, which is initially separated by less than $\eta_k$. (b) Normalized mean-square relative particle-pair dispersion for initial separation of less than $\eta_k$. The inset shows the same data (zoomed) in main figure compensated by the Richardson scaling of $(\tau/\tau_k)^3$.}
\label{fig:two_particle_disp}
\end{figure*}
Pair dispersion quantifies a change in a separation between two tracer particles over a time. Pair dispersion relative to their initial separation at time $t_0$ for a time lag $\tau$ is defined as
\begin{equation}
    \label{eq:R}
    R_i^2(\tau)
    =
    \left\langle
    \left[
    \Delta q_{i,\alpha}(t_0+\tau)-\Delta q_{i,\alpha}(t_0)
    \right]^2
    \right\rangle_{\alpha}.
\end{equation}
$\langle \cdot \rangle_{\alpha}$ is an arithmetic average over $N_{\alpha}$ tracer pairs. We define the lateral contributions as
\begin{equation}
    \label{eq:R_Pdisp}
    R_{xy}^2(\tau) = R_{x}^2(\tau) + R_y^2(\tau),
\end{equation}
and total pair separation as
\begin{equation}
    \label{eq:Rz_Pdisp}
    R^2(\tau) = R_{xy}^2(\tau)+R_z^2(\tau).
\end{equation}
Here, $\alpha$ denotes the particle-pair index, $\Delta q_{i,\alpha}(t)=q_{2,i,\alpha}(t)-q_{1,i,\alpha}(t)$ is the $i$-component of the separation vector between the two particles at time $t$. Note that $N_{\alpha}$ changes with time; we take the average  over all particle pairs that are available at the time lag $t+t_0$ in \eqref{eq:R}. Figure~\ref{fig:two_particle_disp}(a) shows a typical example of a particle pair within the measurement volume with an initial separation smaller than $\eta_k$. 

Figure~\ref{fig:two_particle_disp}(b) presents the mean-square relative dispersion of particle pairs with initial separations smaller than $\eta_k$, including both the lateral and total components. At short time lags, the pairs exhibit ballistic separation, characterised by $R_{x,y}^2\sim\tau^2$ and $R^2\sim\tau^2$. At intermediate time lags, interactions with turbulent vortical structures enhance the separation rate, leading to a superdiffusive, Richardson-like $\tau^3$ scaling regime~\cite{SchumacherPRL2008}. At long time lags, the relative dispersion enters a diffusive regime, with $R_{x,y}^2\sim\tau$ and $R^2\sim\tau$. Owing to vertical confinement and the absence of lateral confinement within the measurement volume, the lateral dispersion is stronger than the vertical dispersion; consequently, the 3D and lateral dispersion curves remain close to each other. The dashed line shows the number of pairs at each time-lag on the right-hand axis, decreasing from over $3\times10^5$ at the smallest time lag to approximately $10^3$ at $\tau/\tau_k=50$. The inset shows the main-panel data compensated by $(\tau/\tau_k)^3$. Interestingly, the compensated curves briefly increase over $10\leq\tau/\tau_k\leq30$, indicating a scaling exponent greater than three in the Richardson-like regime, most clearly for the lateral dispersion. Recently, Ettel et al.~\cite{Ettel2026Lagrangian} found that such super-cubic scaling arises from plume-ejection events within the thermal boundary layers and the associated rapid plume-driven motions. Scaling with exponents larger than 3 were obtained for $Ra\geq 10^8$. To conclude, our experiment reproduces results of former experiment~\cite{Ni_dispersion,Weiss_2024} and simulations~\cite{Schumacher2009} for single particle and particle pair statistics.

\subsection{\label{sec:multi-particle_stats}Multi-particle statistics}
In this section, we first discuss particle-cloud dispersion and deformation. We tracked particle clouds comprising $O(100)$ of particles, with an initial size of $20\eta_k$ and $8\eta_k$, and investigated their dispersion and statistics using a PCA approach similar to that recently employed for initially dissipative-scale particle clouds~\cite{Ettel2026Lagrangian}. Furthermore, we perform spectral analysis of the graph Laplacian and a subsequent $k$-means clustering (an unsupervised machine-learning algorithm) to investigate the splitting and merging of a large particle cloud into distinct clusters.

\begin{figure*}
\centering
{\includegraphics[width=\linewidth,trim=0 0 0 0,clip]{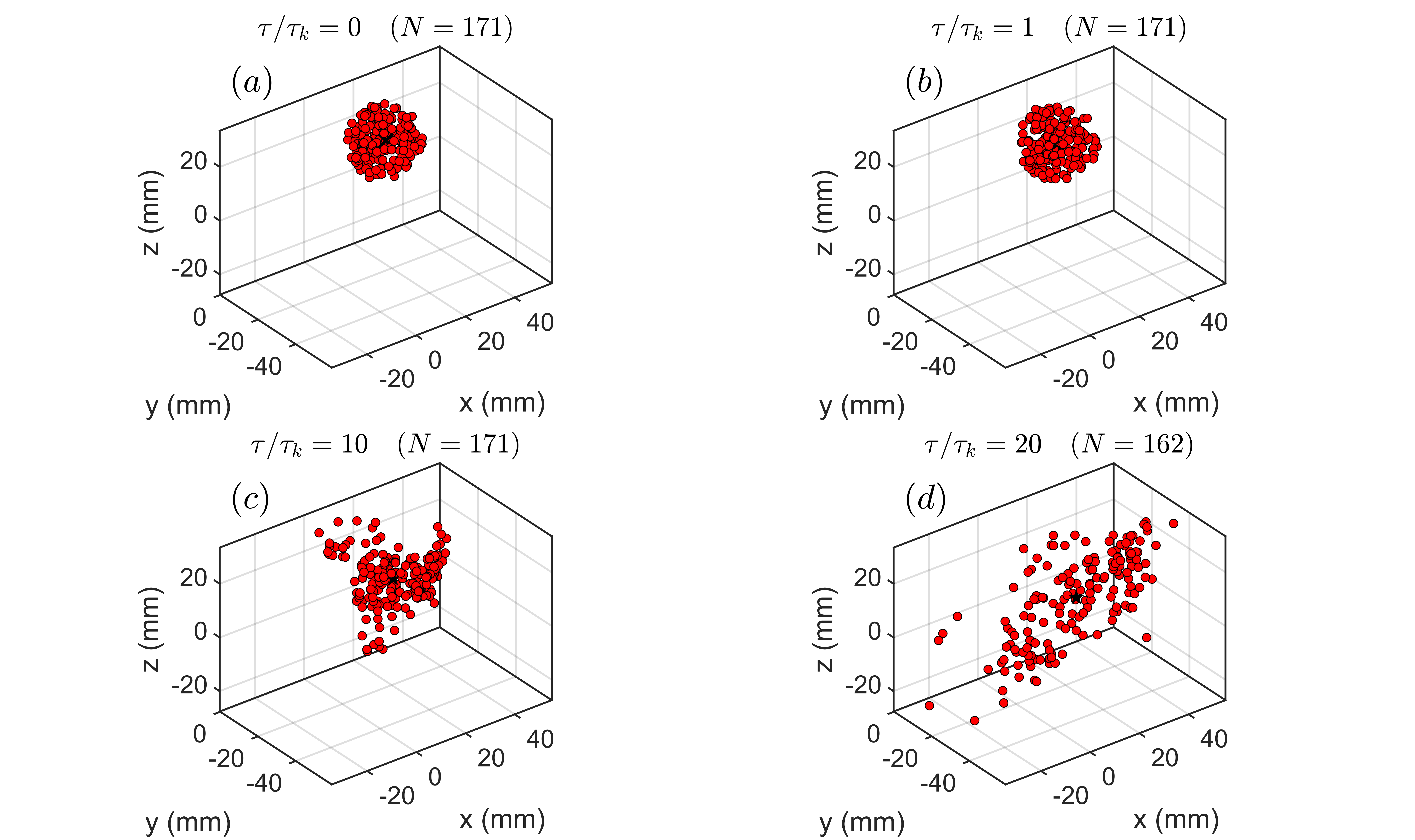}}
\caption{Typical deformation and dispersion of an initially spherical particle cloud: (a) the initial configuration, followed by its evolution into a stretched ellipsoidal shape at (b) $\tau/\tau_k=1$, (c) $\tau/\tau_k=10$, and (d) $\tau/\tau_k=20$. Figures also show a black star marking the centroid of the particles at the time stamp. The sizes of particles are increased for better visualisation.}
\label{fig:swarm}
\end{figure*}
\begin{figure*}
\centering
{\includegraphics[width=0.9\linewidth,trim=0 0 0 0,clip]{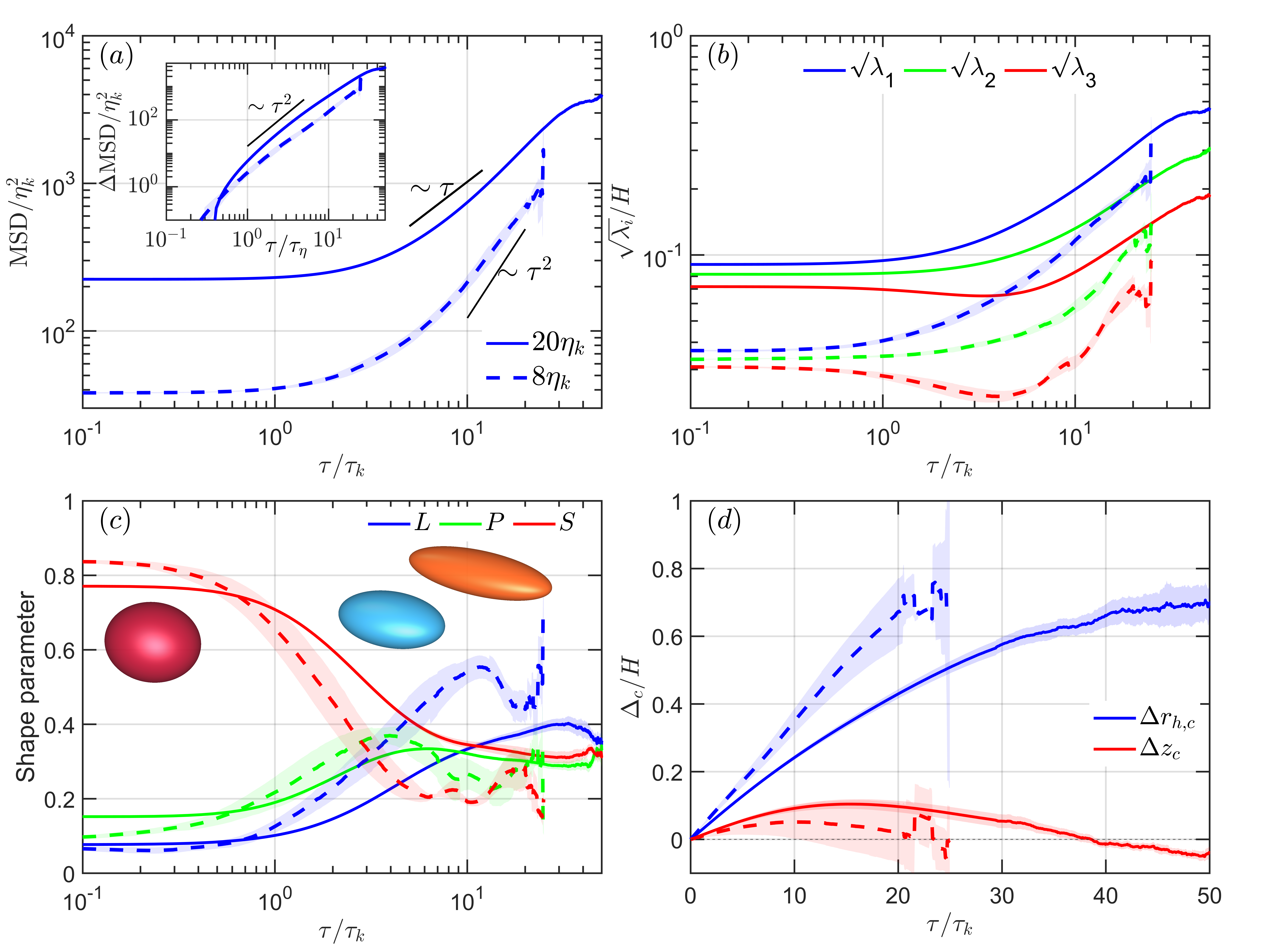}}
\caption{Multi-particle statistics of particle clouds with initial radii of $20\eta_k$ (solid lines) and $8\eta_k$ (dashed lines). (a) Mean-square dispersion of the particle cloud. (b) Maximum, intermediate, and minimum principal spreads as measured by eigenvalues. (c) Shape parameters: linearity $L$ (blue line), planarity $P$ (green line), and sphericity $S$ (red line). Representative mean cloud shapes are shown here: the initial cloud in maroon, and the $20\eta_k$ and $8\eta_k$ particle clouds at later times in bluish and orange colors, respectively. (d) Mean horizontal (blue line) and vertical (red line) motions of the particle-cloud centroid.}
\label{fig:multi-particle}
\end{figure*}
\begin{figure*}
\centering
{\includegraphics[width=\linewidth,trim=0 0 0 0,clip]{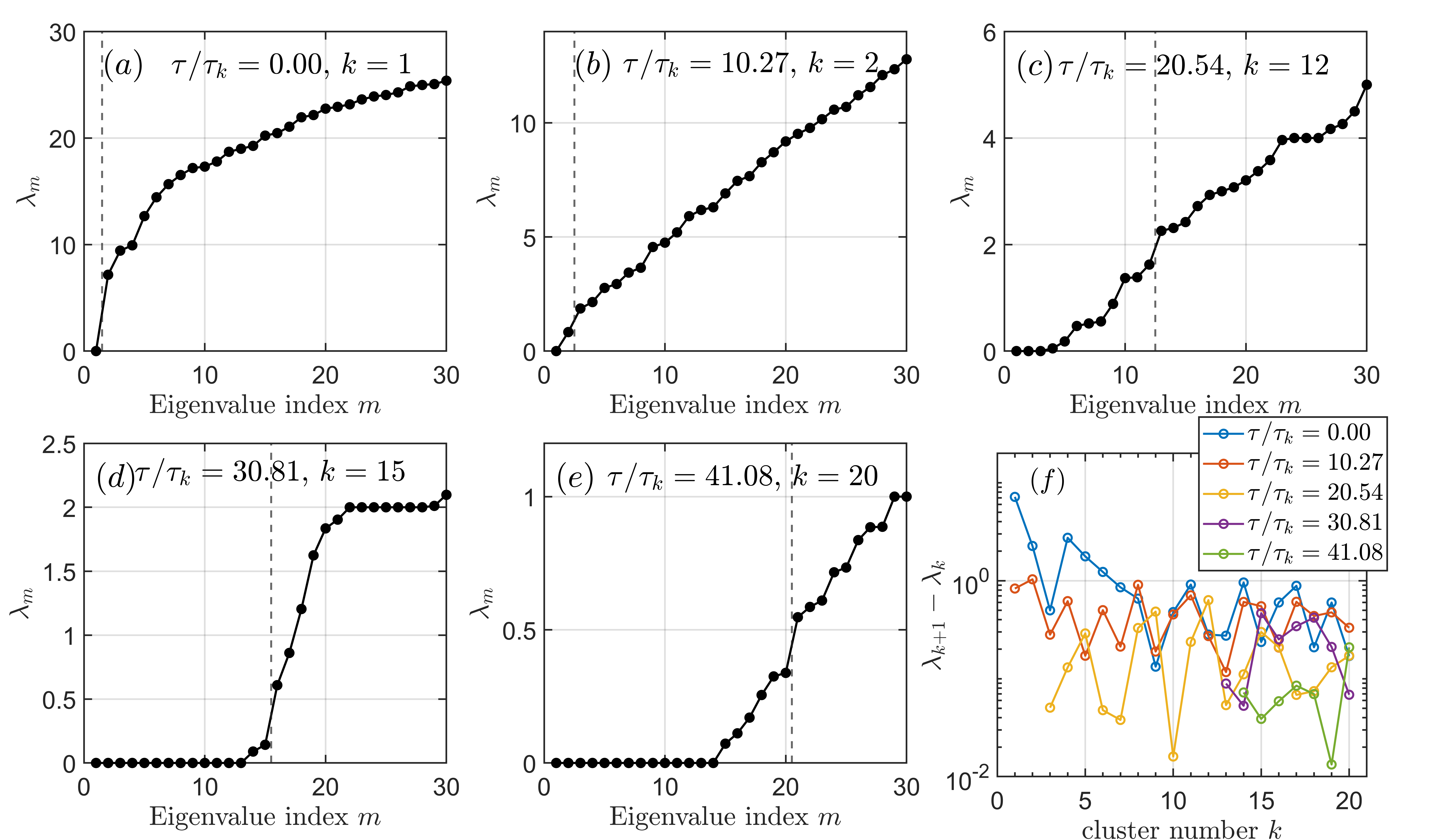}}
\caption{Spectral analysis: Eigenvalue spectra of the graph Laplacian $\hat{L}$ at normalized times $\tau/\tau_k=0$ in (a), $\tau/\tau_k=10.27$ in (b), $\tau/\tau_k=20.54$ in (c), $\tau/\tau_k=30.81$ in (d), and $\tau/\tau_k=41.08$ in (e), together with the corresponding spectral-gap analysis in panel (f). The vertical dashed lines in panels (a)–(e) indicate the maximum spectral gaps, from which $1$, $2$, $12$, $15$, and $20$ particle clusters are identified, respectively. The number of clusters obtained from the spectral analysis is subsequently used as the value of $k$ in the $k$-means clustering algorithm.}
\label{fig:spectral_analysis}
\end{figure*}
\begin{figure*}
\centering
{\includegraphics[width=0.65\linewidth,trim=0.5cm 0 1.5cm 0,clip]{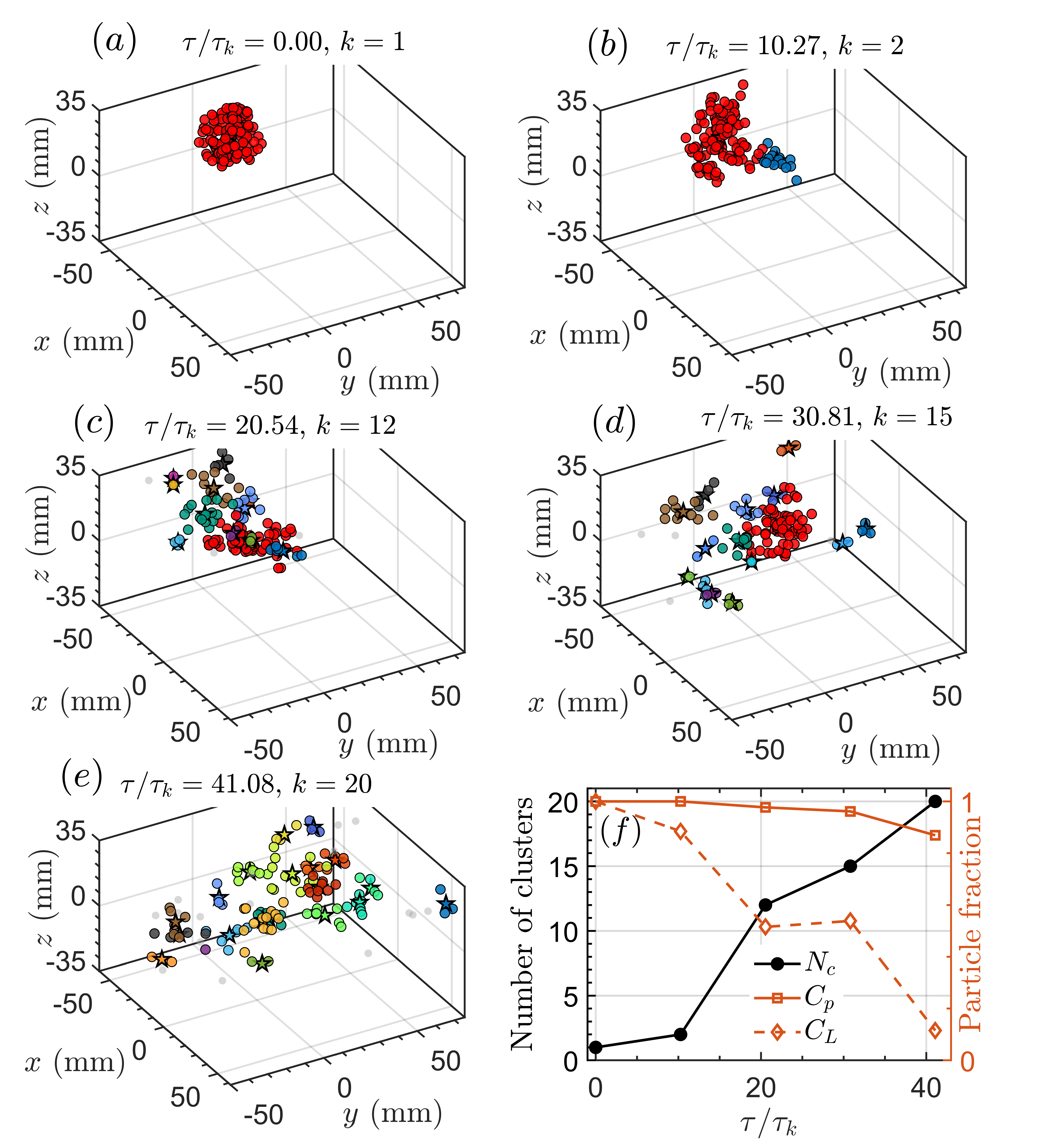}}
\caption{Visualisation of the splitting and merging of the large particle cloud with 130 particles and initial radii of $22\eta_k$ into distinct clusters. The number of identified particle clusters is $1$ at $\tau/\tau_{\eta}=0$ in (a), $2$ at $\tau/\tau_{\eta}=10.27$ in (b), $12$ at $\tau/\tau_{\eta}=20.54$ in (c), $15$ at $\tau/\tau_{\eta}=30.81$ in (d), and $20$ at $\tau/\tau_{\eta}=41.08$ in (e). The corresponding statistical evolution is shown in panel (f), where $N_c$ denotes the number of clusters, $C_p$ the fraction of connected particles, and $C_L$ the particle fraction belonging to the largest cluster. }
\label{fig:clustering}
\end{figure*}

\subsubsection{\label{sec:cloud dispersion}Particle cloud dispersion and deformation}
To better understand pollutant dispersion and accurately identify pollution sources, greater attention is given here to the dispersion and deformation of particle clouds of finite size in laboratory experiments. Figure~\ref{fig:swarm} shows an initially spherical cloud of particles with a radius of $20\eta_k$, for which all particle trajectories are retained up to $10\tau_k$. For the subsequent analysis, we consider only particle clouds for which all constituent particle trajectories are continuously tracked for at least $30\tau_k$. 

At very short times, the particle cloud retains its initial shape. As time progresses, however, the cloud becomes increasingly deformed possibly under the influence of plumes, vortical structures, and the large-scale circulation. Figure~\ref{fig:swarm} shows that the particle cloud, initially located in the upper half of the cell, disperses predominantly in the horizontal direction with time, while also spreading toward the lower half of the cell, possibly due to a downwelling column of fluid in this region. In current analysis, multiple particle-cloud centres are identified within each snapshot, with a minimum separation of $40\eta_k$ between neighbouring centres. A maximum of 50 centres is selected per snapshot, and the selection is performed every 100th snapshot. 

The cloud centres are selected automatically based on the number of neighbouring particles, with regions containing more particles given higher priority. Particle clouds containing fewer than 50 particles are excluded from the analysis because of insufficient statistical sampling. Furthermore, cloud centres are identified separately in the bulk region and near the plates to avoid preferentially selecting centres from only one region. In total, we utilised 1450 particle-cloud centres, with each cloud containing approximately 80--200 particles that were continuously tracked for at least $30\tau_k$. As a second case, particle clouds with an initial radius of $8\eta_k$ were analysed, with all constituent particles continuously tracked for at least $20\tau_k$. In this case, 170 particle-cloud centres were identified, and clouds containing fewer than 25 particles were excluded from the analysis.

Once the particle cloud of $M$ tracers is identified, the gyration tensor can be constructed from the particle positions, defined as
\begin{equation}
\hat{G}=\frac{1}{M-1}\sum_{i=1}^{M}{\bm q}'_i {\bm q}'^{\,T}_i ,
\end{equation}
where ${\bm q}'_i$ the position of the $i$-th Lagrangian particle relative to the particle cloud center. This cloud center given by $\bar{\bm q}=\frac{1}{M}\sum_{i=1}^{M} {\bm q}'_i$; consequently ${\bm q}'_i={\bm q}_i-\bar{\bm q}$. The eigenvalues and eigenvectors are obtained from
\begin{equation}
\hat{G}{\bm v}_k=\lambda_k{\bm v}_k,\qquad\mbox{with}\quad\lambda_1 \geq \lambda_2 \geq \lambda_3 .
\end{equation}
The mean-squared dispersion of the particle cloud is given by
\begin{equation}
\mathrm{MSD_c}=\operatorname{Tr}(\hat{G})=\lambda_1+\lambda_2+\lambda_3 .
\end{equation}
Using the eigenvalues, the shape metrics linearity $L$, planarity $P$, and sphericity $S$ of the particle cloud is given as
\begin{equation}
L=\frac{\lambda_1-\lambda_2}
{\lambda_1+\lambda_2+\lambda_3},
\quad
P=\frac{2(\lambda_2-\lambda_3)}
{\lambda_1+\lambda_2+\lambda_3},
\quad
S=\frac{3\lambda_3}
{\lambda_1+\lambda_2+\lambda_3} \quad\mbox{with}\quad L+P+S=1\,.
\end{equation}
For a spherical particle cloud, the sphericity is $S=1$, corresponding to $\lambda_1\approx\lambda_2\approx\lambda_3$. A high linearity indicates an elongated, tube-like structure, for which $\lambda_1\gg\lambda_2\approx\lambda_3$, whereas a high planarity indicates a sheet-like structure, characterised by $\lambda_1\approx\lambda_2\gg\lambda_3$. 

Figure~\ref{fig:multi-particle} presents the multi-particle dispersion statistics obtained using the PCA approach for particle clouds with initial radii of $20\eta_k$ and $8\eta_k$. Eight independent datasets were used in the present analysis, obtained either from separate, but identical experiments or as statistically independent datasets from the same experiment. In all figures, the mean is calculated across all independent datasets, while the error bands represent the standard error of the mean (SEM) across the datasets. Figure~\ref{fig:multi-particle}(a) shows the mean-square particle separation as a function of time. At very short times, dispersion is negligible, and the particle cloud is transported as a whole with minimal changes in its shape parameters. Beyond the Kolmogorov timescale, however, the dispersion increases rapidly. For $5<\tau/\tau_k<30$, the particle cloud with an initial radius of $20\eta_k$ exhibits neither purely ballistic nor diffusive scaling; instead, an intermediate scaling behaviour is observed. In comparison, the cloud with an initial radius of $8\eta_k$ exhibits a ballistic regime over the same time interval. The principal spreads along the three directions become well separated at longer times, indicating the development of an elongated ellipsoidal particle cloud, as shown in Fig.~\ref{fig:multi-particle}(b). The larger absolute dispersion and principal spreads observed for the particle cloud with an initial radius of $20\eta_k$ are partly a consequence of its larger initial size and should therefore not be interpreted directly as indicating a stronger dispersion rate. We further consider the MSD of the particles belonging to the same cloud relative to its initial value, defined as
\begin{equation}
\Delta \mathrm{MSD}=\mathrm{MSD}-\mathrm{MSD}_0.
\end{equation}
The corresponding results are shown in the inset of Fig.~\ref{fig:multi-particle}(a). For both cloud sizes, the relative-MSD curves show a ballistic regime at short times, extending approximately up to $\tau/\tau_{\eta} \simeq 10$. This behaviour is not evident in the absolute MSD shown in the main panel, likely because of finite-size effects associated with the nonzero initial cloud size. In general, the relative MSD of the larger cloud is greater than that of the smaller cloud. This may be because the larger cloud simultaneously interacts with multiple turbulent eddies and plume structures, thereby enhancing dispersion along several directions, whereas the smaller cloud initially samples a more limited range of flow structures. Thus, within the range of cloud sizes considered here, a larger initial cloud size leads to the greater dispersion.

Figure~\ref{fig:multi-particle}(c) shows the temporal evolution of the linearity $L$, planarity $P$, and sphericity $S$. For the particle cloud with an initial radius of $20\eta_k$, these quantities eventually saturate at values of approximately $1/3$ for $\tau/\tau_k\gtrsim10$, suggesting that, although the particle cloud continues to disperse, its overall shape reaches a statistically stationary state. For the smaller cloud, however, the shape parameters do not saturate at the same value. In particular, the linearity attains a relatively larger value than the planarity and sphericity, indicating a more elongated cloud shape. The mean centroidal displacement in the vertical direction remains close to zero, as expected, whereas the horizontal centroidal displacement increases gradually and saturates for $\tau/\tau_k\gtrsim30$, as shown in Fig.~\ref{fig:multi-particle}(d). The horizontal dispersion of the smaller cloud is somewhat larger than that of the larger cloud at later times. Further, for the larger cloud the angle between the principal axis and the vertical direction increases from $60^{\circ}$ and remains close to $75^\circ$ for longer times while that of the smaller cloud stays close to $60^{\circ}$.  This means that the smaller cloud disperse more in vertical direction relative to the larger cloud. Together with the largest principal spread being 1.5 times greater than the intermediate spread, this indicates preferential dispersion in the horizontal direction. 

This anisotropic dispersion is associated with vertical confinement and the high aspect ratio of the cell. Furthermore, the horizontal dispersion of the larger particle cloud saturates at values smaller than the magnitude of the cell height. Thus, the extent of the particle-cloud dispersion remains much smaller than the half-width of the superstructure (approximately 2.5$H$) at the present $Ra$ and $Pr$~\cite{Pandey_superstructures}. This suggests that the preferential horizontal dispersion occurs predominantly within an individual superstructure roll. This structure can be considered as a nearly coherent set~\cite{Schneide_PRF2018,Vieweg_PRFLetter2021}. In the present analysis, the aspect ratio of the measurement volume is approximately two. Therefore, further studies employing measurement volumes with larger aspect ratios are required to fully determine the magnitude at which the horizontal dispersion saturates.

\subsubsection{\label{sec:clustering}Particle clustering}
As discussed in the previous section, large particle clouds with their larger spatial extent, interact with the turbulent flow as they evolve, resulting in stronger and more anisotropic dispersion. In this subsection, we investigate the formation of smaller particle clusters by tracking the splitting of the bigger particle clouds and the subsequent merging of the resulting sub-clusters.

For the particle-clustering analysis, we select particle clouds of $M$ tracers with an initial size of $22\eta_k$ such that as many particles as possible can be tracked over the longest possible duration. Spectral analysis is performed at five equally spaced time instants, from the initial time up to the latest time for which all particles in the cloud remain tracked. At each time instant, an undirected graph is constructed in which the particles represent the nodes. We choose the minimum connection radius for which all particles initially form a single connected component; this radius is slightly smaller than the initial radius of the particle cloud. Using this fixed connection radius, a binary $M\times M$ adjacency matrix $\hat{W}$ is constructed at each of the five time instants \cite{Schneide2017},
\begin{equation}
    W_{ij}=
\begin{cases}
1, & i\neq j,\\
0, & \mathrm{else}
\end{cases}
\end{equation}
with $W_{ij}=1$ if particles $i$ and $j$ are connected within the chosen radius and $W_{ij}=0$ otherwise. Indices $i,j=1,...,M$. From the resulting adjacency matrix, the diagonal degree matrix $D$ is constructed, with each diagonal element $D_{ii}$ equal to the number of connections of particle $i$ to other particles $j$ (the node degree),
\begin{equation}
 D_{ii}=\sum_{j=1}^{M} W_{ij}\,.  
\end{equation}
The graph Laplacian is defined as $\hat{L}=\hat{D}-\hat{W}$ and the generalized eigenvalue problem 
\begin{equation}
    \hat{L}{\bm v}=\lambda \hat{D}{\bm v}
\end{equation} 
has to be solved. All eigenvalues are real and nonnegative, $0=\lambda_1\leq\lambda_2\leq\cdots\leq\lambda_M$, with corresponding eigenvectors ${\bm v}_1,\ldots, {\bm v}_M$. The ordered eigenvalue spectrum is then examined, and a pronounced gap between eigenvalues $\lambda_n$ and $\lambda_{n+1}$ eigenvalues indicates the presence of $n$ clusters~\cite{Fiedler1973,Luxburg2007,Schneide_PRF2018}. The first $n$ corresponding eigenvectors define a $n$-dimensional spectral representation of the particles, which is subsequently used as the input to the clustering algorithm to assign the particles to the $n$ distinct clusters. Thus, the eigenvalue spectrum determines the number of clusters, whereas the corresponding eigenvectors provide the spectral coordinates derived from the particle-connectivity network, which are used to identify their particle membership.

Figures~\ref{fig:spectral_analysis}(a)--(e) shows the eigenvalue spectra at five different normalized times, together with the corresponding spectral-gap plot in Fig.~\ref{fig:spectral_analysis}(f). The vertical dashed lines in panels (a--e) indicate the locations of the maximum spectral gaps, which subsequently determine the number of clusters for the cluster search algorithm ($k$-means clustering implies the search for $k$ clusters in a data set). As observed, the number of clusters $N_c$ increases with time. Figure~\ref{fig:clustering} visualizes the splitting of the particle cloud and the subsequent formation of distinct clusters. The initially single cloud, shown in red in Fig.\ref{fig:clustering}(a), splits into two clusters of different sizes, shown in red and blue in Fig.~\ref{fig:clustering}(b). The cluster with the maximum overlap with the original cloud retains the color of the parent cloud, whereas the newly formed cluster is assigned a new color. Some particles become disconnected from all clusters and are therefore shown in light grey. The number of disconnected particles is determined directly from the graph connectivity, while the presence of disconnected components is reflected by multiple zero eigenvalues in the ordered eigenvalue spectrum. There are 14 disconnected components at $\tau/\tau_k=41.08$, shown in Figure~\ref{fig:clustering}(e).

In Figs.~\ref{fig:clustering}(b, c), the number of clusters increases rapidly from 2 to 12 over $10<\tau/\tau_k<20$. Subsequently, over $20<\tau/\tau_k<41$, the number of clusters increases further to 20. To this end, we track the number of clusters, $N_c$, the fraction of particles belonging to connected components, $C_p$, and the fraction of particles belonging to the largest cluster, $C_L$. These values are shown in Fig.~\ref{fig:clustering}(f). There may be instances of mergers, where particles previously detached from the parent cloud or other clusters coalesce again to larger clusters while other get disconnected. This is reflected in the increased particle fraction of the largest cluster over $20<\tau/\tau_k<30$.

\begin{figure*}
\centering
{\includegraphics[width=0.9\linewidth,trim=0 0 0 0,clip]{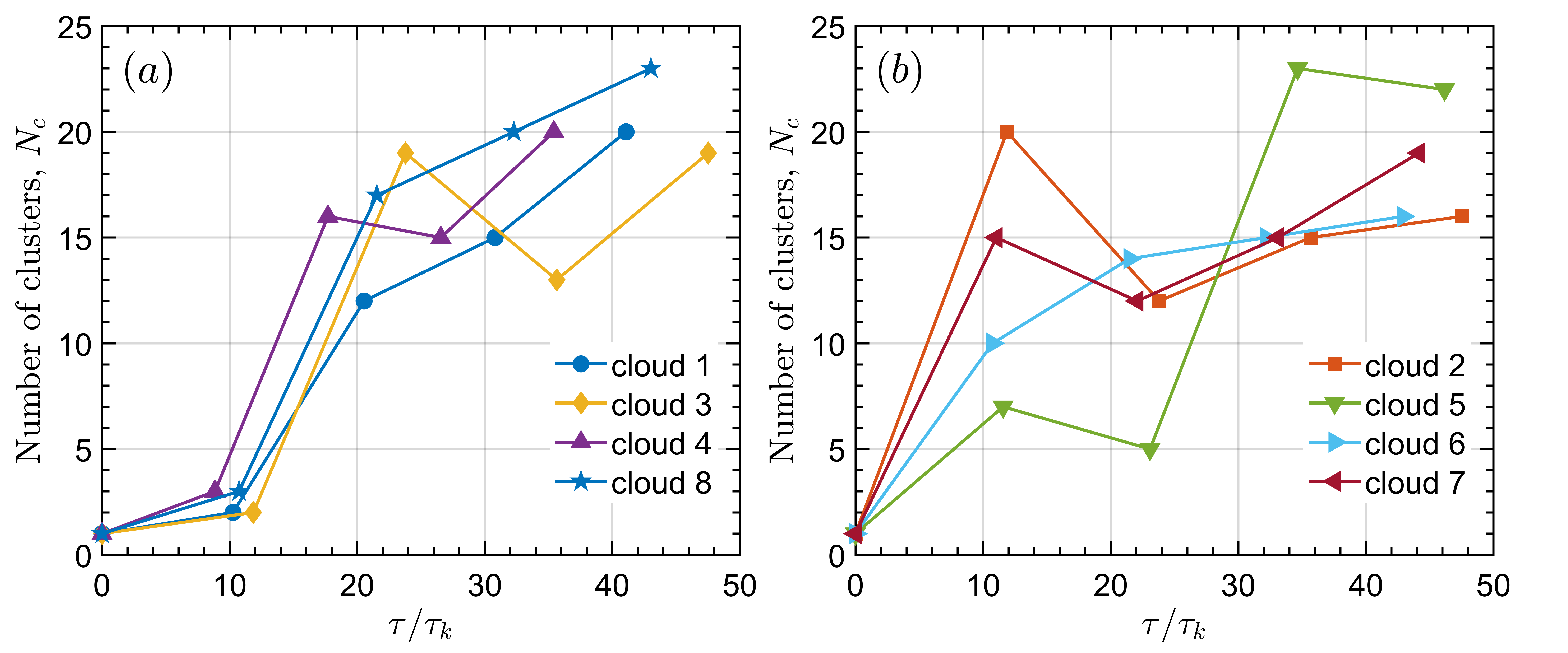}}
\caption{Cluster statistics: The evolution of the number of clusters for eight different cases of particle clouds, grouped into two categories. At early times, $\tau/\tau_k<10$, the clouds show small changes in the number of clusters in (a) while strong changes in the number of clusters are observed in panel (b).}
\label{fig:cluster_stats}
\end{figure*}

Figure~\ref{fig:cluster_stats} shows the evolution of the number of clusters for eight particle clouds, divided into two groups. The first group exhibits a slow increase in the number of clusters up to $\tau/\tau_k=10$, whereas the second group shows a rapid increase over the same time range, as shown in Figures~\ref{fig:cluster_stats}(a) and \ref{fig:cluster_stats}(b), respectively. However, for $10\leq\tau/\tau_k\leq 30$, the number of clusters in the two groups is not substantially different. This may indicate that, at early times, the first group is located in a relatively less turbulent region than the second group. 

To examine this further, we consider the vertical velocity field at mid-height at the initial time. We find that the particle clouds in the first group are located either in upwelling or downwelling regions, as determined by comparing their $(x,y)$ coordinates with the vertical velocity field at mid-height. In contrast, the clouds in the second group span both upwelling and downwelling regions. This suggests that the rapid initial increase in the number of clusters in the second group is associated with the clouds being distributed across differently characterized flow regions. The second group also shows a brief decrease in the number of clusters after $\tau/\tau_k=10$, possibly due to the re-merging of clusters. In both groups, once the number of clusters reaches approximately 15, the rate of increase decreases, possibly because the particle clouds become trapped within the superstructure, which promotes some degree of newly cluster merging.

\section{Conclusions}
We investigated Lagrangian statistics by means of 3D Particle Tracking Velocimetry (PTV) in a turbulent Rayleigh--B\'enard convection cell with an aspect ratio of $\Gamma=10$ using laboratory experiments at $Ra=9.27\times 10^7$ and $Pr=4.84$. Lagrangian particle-position data were obtained using the Shake-the-Box algorithm available in DaVis 10.2. The full depth of the convection cell was illuminated in the central region, sufficiently far from the sidewalls. Particle velocities and accelerations were subsequently obtained from the position data using a penalised B-spline approach.

We presented the probability density functions (PDFs) of all three acceleration components and compared them with corresponding direct numerical simulation (DNS) results. The PDF tails of all acceleration components exhibit stretched-exponential behaviour. The far tails of the vertical acceleration PDF obtained from PTV agree very well with those from DNS, while the lateral components show reasonable agreement within the measurement uncertainty of the PTV data. The high temporal resolution of the measurements provides approximately 50--60 data points within the short timescale over which the acceleration autocorrelation rapidly decays to zero. Well-converged second- and fourth-order moments of the acceleration components are also reported. The lateral acceleration components are found to show a higher intermittency than the vertical component, while the vertical acceleration component exhibits negative skewness, consistent with the DNS results of Schumacher~\cite{SchumacherPRL2008}. Conditional acceleration statistics are further evaluated within a bulk region spanning one-third of the cell height and centred at the mid-height, and are compared with the full-depth measurements. The primary difference between the two regions is that the PDFs in the bulk are generally less intermittent and exhibit lower flatness values than those obtained from the full-depth measurements. This reduction in intermittency is likely due to the exclusion of highly intermittent plume-dominated regions near the plates.

Lagrangian single-particle, particle-pair, and multi-particle statistics were investigated using sufficiently long particle tracks that capture both the short-time ballistic and long-time diffusive regimes. The second-order Lagrangian velocity structure functions transition from short-time ballistic scaling through intermediate scaling regimes to long-time linear scaling at earlier times than the mean-square displacement. This earlier transition reflects the more rapid decorrelation of velocity fluctuations compared with particle displacements and may be associated with the stronger intermittency of the velocity increments. 

Further, the fourth- and sixth-order structure functions are analysed through their relative scaling exponents with respect to the second-order structure function. At short time lags, the exponents exhibit non-intermittent behaviour, while pronounced intermittency emerges as the time lag increases beyond the Kolmogorov time scale, showing trends similar to those reported in refs.\cite{Biferale2008_LVSF,Arneodo2008}. 

The mean-square relative particle-pair dispersion exhibits Batchelor scaling at short times and transitions to a diffusive regime at long times as the particle velocities decorrelate. At intermediate times, we observe Richardson-like scaling with an exponent slightly greater than three, suggesting enhanced relative dispersion, possibly due to buoyancy effects, in agreement with the recent study of Ettel et al.~\cite{Ettel2026Lagrangian}.

The dispersion and deformation of particle clouds with an initial radius of $8\eta_k$ and $20\eta_k$, released from different heights and at different times, were further investigated using a principal component analysis approach. The mean-square dispersion of the particle clouds exhibited a scaling intermediate between ballistic and linear behaviour, possibly owing to their relatively large initial size. In contrast, particle clouds with an initial radius of $8\eta_k$ exhibited a clear ballistic regime. The smaller clouds also showed a larger increase in dispersion and evolved towards a more elongated shape. Furthermore, the major principal axis of the smaller clouds are oriented relatively closer to the vertical direction in relation to that of the larger cloud, indicating a relatively stronger contribution of vertical dispersion. In general, the particle-cloud dispersion was stronger in the lateral directions and remained within the characteristic width of the circulation roll of the superstructure roll. Although the dispersion continued to increase with time, the particle clouds approached a statistically stationary shape. 

Further, owing to the finite spatial extent of the clouds, they interact with plumes and vortices from different regions. At intermediate times, this interaction leads to the formation of distinct particle clusters, while some particles become disconnected. The formation of clusters suggests that particles within a given cluster experience similar local flow conditions or transport processes. In general, the number of clusters increases with time, although some degree of re-merging also occurs.

We are currently working to extend the measurement campaign to simultaneously measure the three-dimensional velocity and temperature fields, enabling the coupled investigation of fluid dynamics and heat transport in high-aspect-ratio turbulent convection experiments. This will require the use of thermochromic liquid crystals as tracer particles, which reflect light at different wavelengths depending on the temperature they experience~\cite{Kaeufer2024Volumetric}.

\acknowledgements
The research of P.P.S. and R.J.S. is funded by the European Union (ERC, MesoComp, 101052786). Views and opinions expressed are, however, those of the authors only and do not necessarily reflect those of the European Union or the European Research Council. The work of M. is supported by a Fellowship of the Free State of Thuringia. The authors would like to thank A. Thieme for support with the experiments. We thank M. Ettel and S. Weiss for helpful discussions.
 
\bibliography{lagn}

\providecommand{\noopsort}[1]{}\providecommand{\singleletter}[1]{#1}%
\begin{thebibliography}{50}%
\makeatletter
\providecommand \@ifxundefined [1]{%
 \@ifx{#1\undefined}
}%
\providecommand \@ifnum [1]{%
 \ifnum #1\expandafter \@firstoftwo
 \else \expandafter \@secondoftwo
 \fi
}%
\providecommand \@ifx [1]{%
 \ifx #1\expandafter \@firstoftwo
 \else \expandafter \@secondoftwo
 \fi
}%
\providecommand \natexlab [1]{#1}%
\providecommand \enquote  [1]{``#1''}%
\providecommand \bibnamefont  [1]{#1}%
\providecommand \bibfnamefont [1]{#1}%
\providecommand \citenamefont [1]{#1}%
\providecommand \href@noop [0]{\@secondoftwo}%
\providecommand \href [0]{\begingroup \@sanitize@url \@href}%
\providecommand \@href[1]{\@@startlink{#1}\@@href}%
\providecommand \@@href[1]{\endgroup#1\@@endlink}%
\providecommand \@sanitize@url [0]{\catcode `\\12\catcode `\$12\catcode `\&12\catcode `\#12\catcode `\^12\catcode `\_12\catcode `\%12\relax}%
\providecommand \@@startlink[1]{}%
\providecommand \@@endlink[0]{}%
\providecommand \url  [0]{\begingroup\@sanitize@url \@url }%
\providecommand \@url [1]{\endgroup\@href {#1}{\urlprefix }}%
\providecommand \urlprefix  [0]{URL }%
\providecommand \Eprint [0]{\href }%
\providecommand \doibase [0]{https://doi.org/}%
\providecommand \selectlanguage [0]{\@gobble}%
\providecommand \bibinfo  [0]{\@secondoftwo}%
\providecommand \bibfield  [0]{\@secondoftwo}%
\providecommand \translation [1]{[#1]}%
\providecommand \BibitemOpen [0]{}%
\providecommand \bibitemStop [0]{}%
\providecommand \bibitemNoStop [0]{.\EOS\space}%
\providecommand \EOS [0]{\spacefactor3000\relax}%
\providecommand \BibitemShut  [1]{\csname bibitem#1\endcsname}%
\let\auto@bib@innerbib\@empty
\bibitem [{\citenamefont {Toschi}\ and\ \citenamefont {Bodenschatz}(2009)}]{toschi_2009_lagrangian}%
  \BibitemOpen
  \bibfield  {author} {\bibinfo {author} {\bibfnamefont {F.}~\bibnamefont {Toschi}}\ and\ \bibinfo {author} {\bibfnamefont {E.}~\bibnamefont {Bodenschatz}},\ }\bibfield  {title} {\bibinfo {title} {Lagrangian properties of particles in turbulence},\ }\href {https://doi.org/10.1146/annurev.fluid.010908.165210} {\bibfield  {journal} {\bibinfo  {journal} {Annu. Rev. Fluid Mech.}\ }\textbf {\bibinfo {volume} {41}},\ \bibinfo {pages} {375} (\bibinfo {year} {2009})}\BibitemShut {NoStop}%
\bibitem [{\citenamefont {Yeung}(2002)}]{Yeung2002Lagrangian}%
  \BibitemOpen
  \bibfield  {author} {\bibinfo {author} {\bibfnamefont {P.~K.}\ \bibnamefont {Yeung}},\ }\bibfield  {title} {\bibinfo {title} {Lagrangian investigations of turbulence},\ }\href {https://doi.org/10.1146/annurev.fluid.34.082101.170725} {\bibfield  {journal} {\bibinfo  {journal} {Annu. Rev. Fluid Mech.}\ }\textbf {\bibinfo {volume} {34}},\ \bibinfo {pages} {115} (\bibinfo {year} {2002})}\BibitemShut {NoStop}%
\bibitem [{\citenamefont {Atkinson}\ and\ \citenamefont {Wu~Zhang}(1996)}]{atkinson}%
  \BibitemOpen
  \bibfield  {author} {\bibinfo {author} {\bibfnamefont {B.~W.}\ \bibnamefont {Atkinson}}\ and\ \bibinfo {author} {\bibfnamefont {J.}~\bibnamefont {Wu~Zhang}},\ }\bibfield  {title} {\bibinfo {title} {Mesoscale shallow convection in the atmosphere},\ }\href {https://doi.org/10.1029/96RG02623} {\bibfield  {journal} {\bibinfo  {journal} {Rev. Geophys.}\ }\textbf {\bibinfo {volume} {34}},\ \bibinfo {pages} {403} (\bibinfo {year} {1996})}\BibitemShut {NoStop}%
\bibitem [{\citenamefont {Pardini}\ \emph {et~al.}(2024)\citenamefont {Pardini}, \citenamefont {Barsotti}, \citenamefont {Bonadonna}, \citenamefont {de' Michieli~Vitturi}, \citenamefont {Folch}, \citenamefont {Mastin}, \citenamefont {Osores},\ and\ \citenamefont {Prata}}]{Pardini2024}%
  \BibitemOpen
  \bibfield  {author} {\bibinfo {author} {\bibfnamefont {F.}~\bibnamefont {Pardini}}, \bibinfo {author} {\bibfnamefont {S.}~\bibnamefont {Barsotti}}, \bibinfo {author} {\bibfnamefont {C.}~\bibnamefont {Bonadonna}}, \bibinfo {author} {\bibfnamefont {M.}~\bibnamefont {de' Michieli~Vitturi}}, \bibinfo {author} {\bibfnamefont {A.}~\bibnamefont {Folch}}, \bibinfo {author} {\bibfnamefont {L.~G.}\ \bibnamefont {Mastin}}, \bibinfo {author} {\bibfnamefont {S.}~\bibnamefont {Osores}},\ and\ \bibinfo {author} {\bibfnamefont {A.~T.}\ \bibnamefont {Prata}},\ }\bibfield  {title} {\bibinfo {title} {Dynamics, monitoring, and forecasting of tephra in the atmosphere},\ }\href {https://doi.org/10.1029/2023RG000808} {\bibfield  {journal} {\bibinfo  {journal} {Rev. Geophys.}\ }\textbf {\bibinfo {volume} {62}},\ \bibinfo {pages} {e2023RG000808} (\bibinfo {year} {2024})}\BibitemShut {NoStop}%
\bibitem [{\citenamefont {DiBenedetto}(2026)}]{DiBenedetto2026}%
  \BibitemOpen
  \bibfield  {author} {\bibinfo {author} {\bibfnamefont {M.~H.}\ \bibnamefont {DiBenedetto}},\ }\bibfield  {title} {\bibinfo {title} {The fluid mechanics of ocean microplastics},\ }\href {https://doi.org/10.1146/annurev-fluid-120423-012604} {\bibfield  {journal} {\bibinfo  {journal} {Annu. Rev. Fluid Mech.}\ }\textbf {\bibinfo {volume} {58}},\ \bibinfo {pages} {355} (\bibinfo {year} {2026})}\BibitemShut {NoStop}%
\bibitem [{\citenamefont {Weiss}\ \emph {et~al.}(2024)\citenamefont {Weiss}, \citenamefont {Schanz}, \citenamefont {Erdogdu}, \citenamefont {Schröder},\ and\ \citenamefont {Bosbach}}]{Weiss_2024}%
  \BibitemOpen
  \bibfield  {author} {\bibinfo {author} {\bibfnamefont {S.}~\bibnamefont {Weiss}}, \bibinfo {author} {\bibfnamefont {D.}~\bibnamefont {Schanz}}, \bibinfo {author} {\bibfnamefont {A.~O.}\ \bibnamefont {Erdogdu}}, \bibinfo {author} {\bibfnamefont {A.}~\bibnamefont {Schröder}},\ and\ \bibinfo {author} {\bibfnamefont {J.}~\bibnamefont {Bosbach}},\ }\bibfield  {title} {\bibinfo {title} {{On Lagrangian properties of turbulent Rayleigh–Bénard convection}},\ }\href {https://doi.org/10.1017/jfm.2024.677} {\bibfield  {journal} {\bibinfo  {journal} {J. Fluid Mech.}\ }\textbf {\bibinfo {volume} {999}},\ \bibinfo {pages} {A90} (\bibinfo {year} {2024})}\BibitemShut {NoStop}%
\bibitem [{\citenamefont {Moller}\ \emph {et~al.}(2022)\citenamefont {Moller}, \citenamefont {Käufer}, \citenamefont {Pandey}, \citenamefont {Schumacher},\ and\ \citenamefont {Cierpka}}]{Moller_Kaufer_Pandey_Schumacher_Cierpka_2022}%
  \BibitemOpen
  \bibfield  {author} {\bibinfo {author} {\bibfnamefont {S.}~\bibnamefont {Moller}}, \bibinfo {author} {\bibfnamefont {T.}~\bibnamefont {Käufer}}, \bibinfo {author} {\bibfnamefont {A.}~\bibnamefont {Pandey}}, \bibinfo {author} {\bibfnamefont {J.}~\bibnamefont {Schumacher}},\ and\ \bibinfo {author} {\bibfnamefont {C.}~\bibnamefont {Cierpka}},\ }\bibfield  {title} {\bibinfo {title} {Combined particle image velocimetry and thermometry of turbulent superstructures in thermal convection},\ }\href {https://doi.org/10.1017/jfm.2022.538} {\bibfield  {journal} {\bibinfo  {journal} {J. Fluid Mech.}\ }\textbf {\bibinfo {volume} {945}},\ \bibinfo {pages} {A22} (\bibinfo {year} {2022})}\BibitemShut {NoStop}%
\bibitem [{\citenamefont {Shevkar}\ \emph {et~al.}(2025{\natexlab{a}})\citenamefont {Shevkar}, \citenamefont {Samuel}, \citenamefont {Cierpka},\ and\ \citenamefont {Schumacher}}]{Shevkar_Samuel_Cierpka_Schumacher_2025}%
  \BibitemOpen
  \bibfield  {author} {\bibinfo {author} {\bibfnamefont {P.~P.}\ \bibnamefont {Shevkar}}, \bibinfo {author} {\bibfnamefont {R.~J.}\ \bibnamefont {Samuel}}, \bibinfo {author} {\bibfnamefont {C.}~\bibnamefont {Cierpka}},\ and\ \bibinfo {author} {\bibfnamefont {J.}~\bibnamefont {Schumacher}},\ }\bibfield  {title} {\bibinfo {title} {Three-dimensional velocity gradient statistics in a mesoscale convection laboratory experiment},\ }\href {https://doi.org/10.1017/jfm.2025.10892} {\bibfield  {journal} {\bibinfo  {journal} {J. Fluid Mech.}\ }\textbf {\bibinfo {volume} {1024}},\ \bibinfo {pages} {A47} (\bibinfo {year} {2025}{\natexlab{a}})}\BibitemShut {NoStop}%
\bibitem [{\citenamefont {Pandey}\ \emph {et~al.}(2018)\citenamefont {Pandey}, \citenamefont {Scheel},\ and\ \citenamefont {P.~Schumacher}}]{Pandey_superstructures}%
  \BibitemOpen
  \bibfield  {author} {\bibinfo {author} {\bibfnamefont {A.}~\bibnamefont {Pandey}}, \bibinfo {author} {\bibfnamefont {J.~D.}\ \bibnamefont {Scheel}},\ and\ \bibinfo {author} {\bibfnamefont {J.}~\bibnamefont {P.~Schumacher}},\ }\bibfield  {title} {\bibinfo {title} {{Turbulent superstructures in Rayleigh-B\'{e}nard convection}},\ }\href {https://doi.org/10.1038/s41467-018-04478-0} {\bibfield  {journal} {\bibinfo  {journal} {Nat. Commun.}\ }\textbf {\bibinfo {volume} {9}},\ \bibinfo {pages} {2118} (\bibinfo {year} {2018})}\BibitemShut {NoStop}%
\bibitem [{\citenamefont {Stevens}\ \emph {et~al.}(2018)\citenamefont {Stevens}, \citenamefont {Blass}, \citenamefont {Zhu}, \citenamefont {Verzicco},\ and\ \citenamefont {Lohse}}]{StevensPRF18}%
  \BibitemOpen
  \bibfield  {author} {\bibinfo {author} {\bibfnamefont {R.~J. A.~M.}\ \bibnamefont {Stevens}}, \bibinfo {author} {\bibfnamefont {A.}~\bibnamefont {Blass}}, \bibinfo {author} {\bibfnamefont {X.}~\bibnamefont {Zhu}}, \bibinfo {author} {\bibfnamefont {R.}~\bibnamefont {Verzicco}},\ and\ \bibinfo {author} {\bibfnamefont {D.}~\bibnamefont {Lohse}},\ }\bibfield  {title} {\bibinfo {title} {{Turbulent thermal superstructures in Rayleigh-Bénard convection}},\ }\href {https://doi.org/10.1103/PhysRevFluids.3.041501} {\bibfield  {journal} {\bibinfo  {journal} {Phys. Rev. Fluids}\ }\textbf {\bibinfo {volume} {3}},\ \bibinfo {pages} {041501(R)} (\bibinfo {year} {2018})}\BibitemShut {NoStop}%
\bibitem [{\citenamefont {Shevkar}\ \emph {et~al.}(2022)\citenamefont {Shevkar}, \citenamefont {Vishnu}, \citenamefont {Mohanan}, \citenamefont {Koothur}, \citenamefont {Mathur},\ and\ \citenamefont {Puthenveettil}}]{shevkar_2022_separating_plumes}%
  \BibitemOpen
  \bibfield  {author} {\bibinfo {author} {\bibfnamefont {P.~P.}\ \bibnamefont {Shevkar}}, \bibinfo {author} {\bibfnamefont {R.}~\bibnamefont {Vishnu}}, \bibinfo {author} {\bibfnamefont {S.~K.}\ \bibnamefont {Mohanan}}, \bibinfo {author} {\bibfnamefont {V.}~\bibnamefont {Koothur}}, \bibinfo {author} {\bibfnamefont {M.}~\bibnamefont {Mathur}},\ and\ \bibinfo {author} {\bibfnamefont {B.~A.}\ \bibnamefont {Puthenveettil}},\ }\bibfield  {title} {\bibinfo {title} {On separating plumes from boundary layers in turbulent convection},\ }\href {https://doi.org/10.1017/jfm.2022.271} {\bibfield  {journal} {\bibinfo  {journal} {J. Fluid Mech.}\ }\textbf {\bibinfo {volume} {941}},\ \bibinfo {pages} {A5} (\bibinfo {year} {2022})}\BibitemShut {NoStop}%
\bibitem [{\citenamefont {Shevkar}\ \emph {et~al.}(2025{\natexlab{b}})\citenamefont {Shevkar}, \citenamefont {Samuel}, \citenamefont {Zinchenko}, \citenamefont {Bode}, \citenamefont {Schumacher},\ and\ \citenamefont {Sreenivasan}}]{Shevkar_etal_2025_hierarchial_network}%
  \BibitemOpen
  \bibfield  {author} {\bibinfo {author} {\bibfnamefont {P.~P.}\ \bibnamefont {Shevkar}}, \bibinfo {author} {\bibfnamefont {R.~J.}\ \bibnamefont {Samuel}}, \bibinfo {author} {\bibfnamefont {G.}~\bibnamefont {Zinchenko}}, \bibinfo {author} {\bibfnamefont {M.}~\bibnamefont {Bode}}, \bibinfo {author} {\bibfnamefont {J.}~\bibnamefont {Schumacher}},\ and\ \bibinfo {author} {\bibfnamefont {K.~R.}\ \bibnamefont {Sreenivasan}},\ }\bibfield  {title} {\bibinfo {title} {{Hierarchical network of thermal plumes and their dynamics in turbulent Rayleigh-Bénard convection}},\ }\href {https://doi.org/10.1073/pnas.2502972122} {\bibfield  {journal} {\bibinfo  {journal} {Proc. Natl. Acad. Sci. USA}\ }\textbf {\bibinfo {volume} {122}},\ \bibinfo {pages} {e2502972122} (\bibinfo {year} {2025}{\natexlab{b}})}\BibitemShut {NoStop}%
\bibitem [{\citenamefont {Sreenivasan}\ and\ \citenamefont {Schumacher}(2010)}]{Sreeni_Joerg_LagnViews}%
  \BibitemOpen
  \bibfield  {author} {\bibinfo {author} {\bibfnamefont {K.~R.}\ \bibnamefont {Sreenivasan}}\ and\ \bibinfo {author} {\bibfnamefont {J.}~\bibnamefont {Schumacher}},\ }\bibfield  {title} {\bibinfo {title} {Lagrangian views on turbulent mixing of passive scalars},\ }\href {https://doi.org/10.1098/rsta.2009.0140} {\bibfield  {journal} {\bibinfo  {journal} {Phil. Trans. R. Soc. A}\ }\textbf {\bibinfo {volume} {368}},\ \bibinfo {pages} {1561} (\bibinfo {year} {2010})}\BibitemShut {NoStop}%
\bibitem [{\citenamefont {Biferale}\ \emph {et~al.}(2008)\citenamefont {Biferale}, \citenamefont {Bodenschatz}, \citenamefont {Cencini}, \citenamefont {Lanotte}, \citenamefont {Ouellette}, \citenamefont {Toschi},\ and\ \citenamefont {Xu}}]{Biferale2008_LVSF}%
  \BibitemOpen
  \bibfield  {author} {\bibinfo {author} {\bibfnamefont {L.}~\bibnamefont {Biferale}}, \bibinfo {author} {\bibfnamefont {E.}~\bibnamefont {Bodenschatz}}, \bibinfo {author} {\bibfnamefont {M.}~\bibnamefont {Cencini}}, \bibinfo {author} {\bibfnamefont {A.~S.}\ \bibnamefont {Lanotte}}, \bibinfo {author} {\bibfnamefont {N.~T.}\ \bibnamefont {Ouellette}}, \bibinfo {author} {\bibfnamefont {F.}~\bibnamefont {Toschi}},\ and\ \bibinfo {author} {\bibfnamefont {H.}~\bibnamefont {Xu}},\ }\bibfield  {title} {\bibinfo {title} {Lagrangian structure functions in turbulence: A quantitative comparison between experiment and direct numerical simulation},\ }\href {https://doi.org/10.1063/1.2930672} {\bibfield  {journal} {\bibinfo  {journal} {Phys. Fluids}\ }\textbf {\bibinfo {volume} {20}},\ \bibinfo {pages} {065103} (\bibinfo {year} {2008})}\BibitemShut {NoStop}%
\bibitem [{\citenamefont {Gasteuil}\ \emph {et~al.}(2007)\citenamefont {Gasteuil}, \citenamefont {Shew}, \citenamefont {Gibert}, \citenamefont {Chill{\'a}}, \citenamefont {Castaing},\ and\ \citenamefont {Pinton}}]{Gasteuil2007}%
  \BibitemOpen
  \bibfield  {author} {\bibinfo {author} {\bibfnamefont {Y.}~\bibnamefont {Gasteuil}}, \bibinfo {author} {\bibfnamefont {W.~L.}\ \bibnamefont {Shew}}, \bibinfo {author} {\bibfnamefont {M.}~\bibnamefont {Gibert}}, \bibinfo {author} {\bibfnamefont {F.}~\bibnamefont {Chill{\'a}}}, \bibinfo {author} {\bibfnamefont {B.}~\bibnamefont {Castaing}},\ and\ \bibinfo {author} {\bibfnamefont {J.-F.}\ \bibnamefont {Pinton}},\ }\bibfield  {title} {\bibinfo {title} {Lagrangian temperature, velocity, and local heat flux measurement in {Rayleigh--B{\'e}nard convection}},\ }\href {https://doi.org/10.1103/PhysRevLett.99.234302} {\bibfield  {journal} {\bibinfo  {journal} {Phys. Rev. Lett.}\ }\textbf {\bibinfo {volume} {99}},\ \bibinfo {pages} {234302} (\bibinfo {year} {2007})}\BibitemShut {NoStop}%
\bibitem [{\citenamefont {Ni}\ \emph {et~al.}(2012)\citenamefont {Ni}, \citenamefont {Huang},\ and\ \citenamefont {Xia}}]{Ni_Huang_Xia_2012}%
  \BibitemOpen
  \bibfield  {author} {\bibinfo {author} {\bibfnamefont {R.}~\bibnamefont {Ni}}, \bibinfo {author} {\bibfnamefont {S.}~\bibnamefont {Huang}},\ and\ \bibinfo {author} {\bibfnamefont {K.}~\bibnamefont {Xia}},\ }\bibfield  {title} {\bibinfo {title} {Lagrangian acceleration measurements in convective thermal turbulence},\ }\href {https://doi.org/10.1017/jfm.2011.520} {\bibfield  {journal} {\bibinfo  {journal} {J. Fluid Mech.}\ }\textbf {\bibinfo {volume} {692}},\ \bibinfo {pages} {395–419} (\bibinfo {year} {2012})}\BibitemShut {NoStop}%
\bibitem [{\citenamefont {Shnapp}\ and\ \citenamefont {Liberzon}(2018)}]{Shnapp_Liberzon_2018}%
  \BibitemOpen
  \bibfield  {author} {\bibinfo {author} {\bibfnamefont {R.}~\bibnamefont {Shnapp}}\ and\ \bibinfo {author} {\bibfnamefont {A.}~\bibnamefont {Liberzon}},\ }\bibfield  {title} {\bibinfo {title} {Generalization of turbulent pair dispersion to large initial separations},\ }\href {https://doi.org/10.1103/PhysRevLett.120.244502} {\bibfield  {journal} {\bibinfo  {journal} {Phys. Rev. Lett.}\ }\textbf {\bibinfo {volume} {120}},\ \bibinfo {pages} {244502} (\bibinfo {year} {2018})}\BibitemShut {NoStop}%
\bibitem [{\citenamefont {Mathai}\ \emph {et~al.}(2018)\citenamefont {Mathai}, \citenamefont {Huisman}, \citenamefont {Sun}, \citenamefont {Lohse},\ and\ \citenamefont {Bourgoin}}]{Varghese2018}%
  \BibitemOpen
  \bibfield  {author} {\bibinfo {author} {\bibfnamefont {V.}~\bibnamefont {Mathai}}, \bibinfo {author} {\bibfnamefont {S.~G.}\ \bibnamefont {Huisman}}, \bibinfo {author} {\bibfnamefont {C.}~\bibnamefont {Sun}}, \bibinfo {author} {\bibfnamefont {D.}~\bibnamefont {Lohse}},\ and\ \bibinfo {author} {\bibfnamefont {M.}~\bibnamefont {Bourgoin}},\ }\bibfield  {title} {\bibinfo {title} {Dispersion of air bubbles in isotropic turbulence},\ }\href {https://doi.org/10.1103/PhysRevLett.121.054501} {\bibfield  {journal} {\bibinfo  {journal} {Phys. Rev. Lett.}\ }\textbf {\bibinfo {volume} {121}},\ \bibinfo {pages} {054501} (\bibinfo {year} {2018})}\BibitemShut {NoStop}%
\bibitem [{\citenamefont {Wang}\ \emph {et~al.}(2025)\citenamefont {Wang}, \citenamefont {Huisman}, \citenamefont {Basset}, \citenamefont {Volk},\ and\ \citenamefont {Bourgoin}}]{Wang_HIT2026}%
  \BibitemOpen
  \bibfield  {author} {\bibinfo {author} {\bibfnamefont {C.}~\bibnamefont {Wang}}, \bibinfo {author} {\bibfnamefont {S.~G.}\ \bibnamefont {Huisman}}, \bibinfo {author} {\bibfnamefont {T.}~\bibnamefont {Basset}}, \bibinfo {author} {\bibfnamefont {R.}~\bibnamefont {Volk}},\ and\ \bibinfo {author} {\bibfnamefont {M.}~\bibnamefont {Bourgoin}},\ }\bibfield  {title} {\bibinfo {title} {Lagrangian flow statistics in experimental homogeneous isotropic turbulence},\ }\href {https://doi.org/10.1063/5.0267921} {\bibfield  {journal} {\bibinfo  {journal} {Phys. Fluids}\ }\textbf {\bibinfo {volume} {37}},\ \bibinfo {pages} {075164} (\bibinfo {year} {2025})}\BibitemShut {NoStop}%
\bibitem [{\citenamefont {Li}\ \emph {et~al.}(2026)\citenamefont {Li}, \citenamefont {Zhang}, \citenamefont {Fan}, \citenamefont {Toschi},\ and\ \citenamefont {Sun}}]{Li2026LagrangianDroplets}%
  \BibitemOpen
  \bibfield  {author} {\bibinfo {author} {\bibfnamefont {L.}~\bibnamefont {Li}}, \bibinfo {author} {\bibfnamefont {Y.-B.}\ \bibnamefont {Zhang}}, \bibinfo {author} {\bibfnamefont {Y.}~\bibnamefont {Fan}}, \bibinfo {author} {\bibfnamefont {F.}~\bibnamefont {Toschi}},\ and\ \bibinfo {author} {\bibfnamefont {C.}~\bibnamefont {Sun}},\ }\bibfield  {title} {\bibinfo {title} {{Experimental investigation into Lagrangian statistics of droplets in homogeneous isotropic turbulence}},\ }\href {https://doi.org/10.1209/0295-5075/ae41e3} {\bibfield  {journal} {\bibinfo  {journal} {Europhys. Lett.}\ }\textbf {\bibinfo {volume} {153}},\ \bibinfo {pages} {43001} (\bibinfo {year} {2026})}\BibitemShut {NoStop}%
\bibitem [{\citenamefont {Emran}\ and\ \citenamefont {Schumacher}(2010)}]{EmranSchumacher2010_TracerDy}%
  \BibitemOpen
  \bibfield  {author} {\bibinfo {author} {\bibfnamefont {M.~S.}\ \bibnamefont {Emran}}\ and\ \bibinfo {author} {\bibfnamefont {J.}~\bibnamefont {Schumacher}},\ }\bibfield  {title} {\bibinfo {title} {Lagrangian tracer dynamics in a closed cylindrical turbulent convection cell},\ }\href {https://doi.org/10.1103/PhysRevE.82.016303} {\bibfield  {journal} {\bibinfo  {journal} {Phys. Rev. E}\ }\textbf {\bibinfo {volume} {82}},\ \bibinfo {pages} {016303} (\bibinfo {year} {2010})}\BibitemShut {NoStop}%
\bibitem [{\citenamefont {Ni}\ and\ \citenamefont {Xia}(2013)}]{Ni_dispersion}%
  \BibitemOpen
  \bibfield  {author} {\bibinfo {author} {\bibfnamefont {R.}~\bibnamefont {Ni}}\ and\ \bibinfo {author} {\bibfnamefont {K.~Q.}\ \bibnamefont {Xia}},\ }\bibfield  {title} {\bibinfo {title} {Experimental investigation of pair dispersion with small initial separation in convective turbulent flows},\ }\href {https://doi.org/10.1103/PhysRevE.87.063006} {\bibfield  {journal} {\bibinfo  {journal} {Phys. Rev. E}\ }\textbf {\bibinfo {volume} {87}},\ \bibinfo {pages} {063006} (\bibinfo {year} {2013})}\BibitemShut {NoStop}%
\bibitem [{\citenamefont {Li}\ \emph {et~al.}(2021)\citenamefont {Li}, \citenamefont {Huang}, \citenamefont {Ni},\ and\ \citenamefont {Xia}}]{Li_Huang_Ni_Xia_2021}%
  \BibitemOpen
  \bibfield  {author} {\bibinfo {author} {\bibfnamefont {X.-M.}\ \bibnamefont {Li}}, \bibinfo {author} {\bibfnamefont {S.-D.}\ \bibnamefont {Huang}}, \bibinfo {author} {\bibfnamefont {R.}~\bibnamefont {Ni}},\ and\ \bibinfo {author} {\bibfnamefont {K.-Q.}\ \bibnamefont {Xia}},\ }\bibfield  {title} {\bibinfo {title} {Lagrangian velocity and acceleration measurements in plume-rich regions of turbulent rayleigh--b{\'e}nard convection},\ }\href {https://doi.org/10.1103/PhysRevFluids.6.053503} {\bibfield  {journal} {\bibinfo  {journal} {Phys. Rev. Fluids}\ }\textbf {\bibinfo {volume} {6}},\ \bibinfo {pages} {053503} (\bibinfo {year} {2021})}\BibitemShut {NoStop}%
\bibitem [{\citenamefont {Schumacher}(2009)}]{Schumacher2009}%
  \BibitemOpen
  \bibfield  {author} {\bibinfo {author} {\bibfnamefont {J.}~\bibnamefont {Schumacher}},\ }\bibfield  {title} {\bibinfo {title} {Lagrangian studies in convective turbulence},\ }\href {https://doi.org/10.1103/PhysRevE.79.056301} {\bibfield  {journal} {\bibinfo  {journal} {Phys. Rev. E}\ }\textbf {\bibinfo {volume} {79}},\ \bibinfo {pages} {056301} (\bibinfo {year} {2009})}\BibitemShut {NoStop}%
\bibitem [{\citenamefont {Schanz}\ \emph {et~al.}(2016)\citenamefont {Schanz}, \citenamefont {Gesemann},\ and\ \citenamefont {Schröder}}]{schanz_2016_shakethebox}%
  \BibitemOpen
  \bibfield  {author} {\bibinfo {author} {\bibfnamefont {D.}~\bibnamefont {Schanz}}, \bibinfo {author} {\bibfnamefont {S.}~\bibnamefont {Gesemann}},\ and\ \bibinfo {author} {\bibfnamefont {A.}~\bibnamefont {Schröder}},\ }\bibfield  {title} {\bibinfo {title} {{Shake-The-Box: Lagrangian particle tracking at high particle image densities}},\ }\href {https://doi.org/10.1007/s00348-016-2157-1} {\bibfield  {journal} {\bibinfo  {journal} {Exp. Fluids}\ }\textbf {\bibinfo {volume} {57}},\ \bibinfo {pages} {70} (\bibinfo {year} {2016})}\BibitemShut {NoStop}%
\bibitem [{\citenamefont {K{\"a}ufer}\ \emph {et~al.}(2023)\citenamefont {K{\"a}ufer}, \citenamefont {Vieweg}, \citenamefont {Schumacher},\ and\ \citenamefont {Cierpka}}]{Kaeufer2023}%
  \BibitemOpen
  \bibfield  {author} {\bibinfo {author} {\bibfnamefont {T.}~\bibnamefont {K{\"a}ufer}}, \bibinfo {author} {\bibfnamefont {P.~P.}\ \bibnamefont {Vieweg}}, \bibinfo {author} {\bibfnamefont {J.}~\bibnamefont {Schumacher}},\ and\ \bibinfo {author} {\bibfnamefont {C.}~\bibnamefont {Cierpka}},\ }\bibfield  {title} {\bibinfo {title} {Thermal boundary condition studies in large aspect ratio rayleigh--b{\'e}nard convection},\ }\href {https://doi.org/10.1016/j.euromechflu.2023.06.003} {\bibfield  {journal} {\bibinfo  {journal} {Eur. J. Mech. - B/Fluids}\ }\textbf {\bibinfo {volume} {101}},\ \bibinfo {pages} {283} (\bibinfo {year} {2023})}\BibitemShut {NoStop}%
\bibitem [{\citenamefont {Vieweg}\ \emph {et~al.}(2025)\citenamefont {Vieweg}, \citenamefont {Käufer}, \citenamefont {Cierpka},\ and\ \citenamefont {Schumacher}}]{Vieweg2025}%
  \BibitemOpen
  \bibfield  {author} {\bibinfo {author} {\bibfnamefont {P.~P.}\ \bibnamefont {Vieweg}}, \bibinfo {author} {\bibfnamefont {T.}~\bibnamefont {Käufer}}, \bibinfo {author} {\bibfnamefont {C.}~\bibnamefont {Cierpka}},\ and\ \bibinfo {author} {\bibfnamefont {J.}~\bibnamefont {Schumacher}},\ }\bibfield  {title} {\bibinfo {title} {{Digital twin of a large-aspect-ratio Rayleigh–B\'{e}nard experiment: role of thermal boundary conditions, measurement errors and uncertainties}},\ }\href {https://doi.org/10.1017/flo.2024.35} {\bibfield  {journal} {\bibinfo  {journal} {Flow}\ }\textbf {\bibinfo {volume} {5}},\ \bibinfo {pages} {{E4}} (\bibinfo {year} {2025})}\BibitemShut {NoStop}%
\bibitem [{\citenamefont {Grossmann}\ and\ \citenamefont {Lohse}(2000)}]{GROSSMANN_LOHSE_2000}%
  \BibitemOpen
  \bibfield  {author} {\bibinfo {author} {\bibfnamefont {S.}~\bibnamefont {Grossmann}}\ and\ \bibinfo {author} {\bibfnamefont {D.}~\bibnamefont {Lohse}},\ }\bibfield  {title} {\bibinfo {title} {Scaling in thermal convection: a unifying theory},\ }\href {https://doi.org/10.1017/S0022112099007545} {\bibfield  {journal} {\bibinfo  {journal} {J. Fluid Mech.}\ }\textbf {\bibinfo {volume} {407}},\ \bibinfo {pages} {27} (\bibinfo {year} {2000})}\BibitemShut {NoStop}%
\bibitem [{\citenamefont {Wieneke}(2008)}]{wieneke2008}%
  \BibitemOpen
  \bibfield  {author} {\bibinfo {author} {\bibfnamefont {B.}~\bibnamefont {Wieneke}},\ }\bibfield  {title} {\bibinfo {title} {Volume self-calibration for 3d particle image velocimetry},\ }\href {https://doi.org/10.1007/s00348-008-0521-5} {\bibfield  {journal} {\bibinfo  {journal} {Exp. Fluids}\ }\textbf {\bibinfo {volume} {45}},\ \bibinfo {pages} {549} (\bibinfo {year} {2008})}\BibitemShut {NoStop}%
\bibitem [{\citenamefont {Schanz}\ \emph {et~al.}(2012)\citenamefont {Schanz}, \citenamefont {Gesemann}, \citenamefont {Schröder}, \citenamefont {Wieneke},\ and\ \citenamefont {Novara}}]{Schanz2012_OTF}%
  \BibitemOpen
  \bibfield  {author} {\bibinfo {author} {\bibfnamefont {D.}~\bibnamefont {Schanz}}, \bibinfo {author} {\bibfnamefont {S.}~\bibnamefont {Gesemann}}, \bibinfo {author} {\bibfnamefont {A.}~\bibnamefont {Schröder}}, \bibinfo {author} {\bibfnamefont {B.}~\bibnamefont {Wieneke}},\ and\ \bibinfo {author} {\bibfnamefont {M.}~\bibnamefont {Novara}},\ }\bibfield  {title} {\bibinfo {title} {Non-uniform optical transfer functions in particle imaging: Calibration and application to tomographic reconstruction},\ }\href {https://doi.org/10.1088/0957-0233/24/2/024009} {\bibfield  {journal} {\bibinfo  {journal} {Meas. Sci. Technol.}\ }\textbf {\bibinfo {volume} {24}},\ \bibinfo {pages} {024009} (\bibinfo {year} {2012})}\BibitemShut {NoStop}%
\bibitem [{\citenamefont {Schanz}\ \emph {et~al.}(2021)\citenamefont {Schanz}, \citenamefont {Novara},\ and\ \citenamefont {Schröder}}]{Schanz2021_VTSTB}%
  \BibitemOpen
  \bibfield  {author} {\bibinfo {author} {\bibfnamefont {D.}~\bibnamefont {Schanz}}, \bibinfo {author} {\bibfnamefont {M.}~\bibnamefont {Novara}},\ and\ \bibinfo {author} {\bibfnamefont {A.}~\bibnamefont {Schröder}},\ }\bibfield  {title} {\bibinfo {title} {Shake-the-box particle tracking with variable time-steps in flows with high velocity range (vt-stb)},\ }in\ \href {https://doi.org/10.18409/ispiv.v1i1.65} {\emph {\bibinfo {booktitle} {Proceedings of the 14th International Symposium on Particle Image Velocimetry}}}\ (\bibinfo {address} {Chicago, IL, USA},\ \bibinfo {year} {2021})\BibitemShut {NoStop}%
\bibitem [{\citenamefont {Eilers}\ and\ \citenamefont {Marx}(1996)}]{Eilers1996PSplines}%
  \BibitemOpen
  \bibfield  {author} {\bibinfo {author} {\bibfnamefont {P.~H.~C.}\ \bibnamefont {Eilers}}\ and\ \bibinfo {author} {\bibfnamefont {B.~D.}\ \bibnamefont {Marx}},\ }\bibfield  {title} {\bibinfo {title} {Flexible smoothing with {B}-splines and penalties},\ }\href {https://doi.org/10.1214/ss/1038425655} {\bibfield  {journal} {\bibinfo  {journal} {Stat. Sci.}\ }\textbf {\bibinfo {volume} {11}},\ \bibinfo {pages} {89} (\bibinfo {year} {1996})}\BibitemShut {NoStop}%
\bibitem [{\citenamefont {Gesemann}\ \emph {et~al.}(2016)\citenamefont {Gesemann}, \citenamefont {Huhn}, \citenamefont {Schanz},\ and\ \citenamefont {Schr{\"o}der}}]{Gesemann2016NoisyParticleTracks}%
  \BibitemOpen
  \bibfield  {author} {\bibinfo {author} {\bibfnamefont {S.}~\bibnamefont {Gesemann}}, \bibinfo {author} {\bibfnamefont {F.}~\bibnamefont {Huhn}}, \bibinfo {author} {\bibfnamefont {D.}~\bibnamefont {Schanz}},\ and\ \bibinfo {author} {\bibfnamefont {A.}~\bibnamefont {Schr{\"o}der}},\ }\bibfield  {title} {\bibinfo {title} {From noisy particle tracks to velocity, acceleration and pressure fields using b-splines and penalties},\ }in\ \href@noop {} {\emph {\bibinfo {booktitle} {Proceedings of the 18th International Symposium on Applications of Laser and Imaging Techniques to Fluid Mechanics}}}\ (\bibinfo {address} {Lisbon, Portugal},\ \bibinfo {year} {2016})\BibitemShut {NoStop}%
\bibitem [{\citenamefont {Cheminet}\ \emph {et~al.}(2021)\citenamefont {Cheminet}, \citenamefont {Ostovan}, \citenamefont {Valori}, \citenamefont {Cuvier}, \citenamefont {Daviaud}, \citenamefont {Debue}, \citenamefont {Dubrulle}, \citenamefont {Foucaut},\ and\ \citenamefont {Laval}}]{Cheminet2021RegularizedBSplineSmoothing}%
  \BibitemOpen
  \bibfield  {author} {\bibinfo {author} {\bibfnamefont {A.}~\bibnamefont {Cheminet}}, \bibinfo {author} {\bibfnamefont {Y.}~\bibnamefont {Ostovan}}, \bibinfo {author} {\bibfnamefont {V.}~\bibnamefont {Valori}}, \bibinfo {author} {\bibfnamefont {C.}~\bibnamefont {Cuvier}}, \bibinfo {author} {\bibfnamefont {F.}~\bibnamefont {Daviaud}}, \bibinfo {author} {\bibfnamefont {P.}~\bibnamefont {Debue}}, \bibinfo {author} {\bibfnamefont {B.}~\bibnamefont {Dubrulle}}, \bibinfo {author} {\bibfnamefont {J.-M.}\ \bibnamefont {Foucaut}},\ and\ \bibinfo {author} {\bibfnamefont {J.-P.}\ \bibnamefont {Laval}},\ }\bibfield  {title} {\bibinfo {title} {Optimization of regularized b-spline smoothing for turbulent lagrangian trajectories},\ }\href {https://doi.org/10.1016/j.expthermflusci.2021.110376} {\bibfield  {journal} {\bibinfo  {journal} {Exp. Thermal Fluid Sci.}\ }\textbf {\bibinfo {volume} {127}},\ \bibinfo {pages} {110376} (\bibinfo {year} {2021})}\BibitemShut {NoStop}%
\bibitem [{\citenamefont {Schumacher}(2008)}]{SchumacherPRL2008}%
  \BibitemOpen
  \bibfield  {author} {\bibinfo {author} {\bibfnamefont {J.}~\bibnamefont {Schumacher}},\ }\bibfield  {title} {\bibinfo {title} {Lagrangian dispersion and heat transport in convective turbulence},\ }\href {https://doi.org/10.1103/PhysRevLett.100.134502} {\bibfield  {journal} {\bibinfo  {journal} {Phys. Rev. Lett.}\ }\textbf {\bibinfo {volume} {100}},\ \bibinfo {pages} {134502} (\bibinfo {year} {2008})}\BibitemShut {NoStop}%
\bibitem [{\citenamefont {Sciacchitano}\ \emph {et~al.}(2026)\citenamefont {Sciacchitano}, \citenamefont {Godbersen}, \citenamefont {Grauer}, \citenamefont {Grille~Guerra}, \citenamefont {Leclaire}, \citenamefont {Schanz},\ and\ \citenamefont {Schr{\"o}der}}]{Sciacchitano2026SecondLPTChallenge}%
  \BibitemOpen
  \bibfield  {author} {\bibinfo {author} {\bibfnamefont {A.}~\bibnamefont {Sciacchitano}}, \bibinfo {author} {\bibfnamefont {P.}~\bibnamefont {Godbersen}}, \bibinfo {author} {\bibfnamefont {S.}~\bibnamefont {Grauer}}, \bibinfo {author} {\bibfnamefont {A.}~\bibnamefont {Grille~Guerra}}, \bibinfo {author} {\bibfnamefont {B.}~\bibnamefont {Leclaire}}, \bibinfo {author} {\bibfnamefont {D.}~\bibnamefont {Schanz}},\ and\ \bibinfo {author} {\bibfnamefont {A.}~\bibnamefont {Schr{\"o}der}},\ }\bibfield  {title} {\bibinfo {title} {Main results of the second {LPT} challenge},\ }in\ \href@noop {} {\emph {\bibinfo {booktitle} {Proceedings of the 22nd International Symposium on Applications of Laser and Imaging Techniques to Fluid Mechanics}}}\ (\bibinfo {year} {2026})\BibitemShut {NoStop}%
\bibitem [{\citenamefont {Taylor}(1922)}]{taylor_1922_diffusion}%
  \BibitemOpen
  \bibfield  {author} {\bibinfo {author} {\bibfnamefont {G.~I.}\ \bibnamefont {Taylor}},\ }\bibfield  {title} {\bibinfo {title} {Diffusion by continuous movements},\ }\href {https://doi.org/10.1112/plms/s2-20.1.196} {\bibfield  {journal} {\bibinfo  {journal} {Proc. Lond. Math. Soc.}\ }\textbf {\bibinfo {volume} {s2-20}},\ \bibinfo {pages} {196} (\bibinfo {year} {1922})}\BibitemShut {NoStop}%
\bibitem [{\citenamefont {Richardson}(1926)}]{richardson_1926_atmospheric}%
  \BibitemOpen
  \bibfield  {author} {\bibinfo {author} {\bibfnamefont {L.~F.}\ \bibnamefont {Richardson}},\ }\bibfield  {title} {\bibinfo {title} {Atmospheric diffusion shown on a distance-neighbour graph},\ }\href {https://doi.org/10.1098/rspa.1926.0043} {\bibfield  {journal} {\bibinfo  {journal} {Proc. R. Soc. Lond. A}\ }\textbf {\bibinfo {volume} {110}},\ \bibinfo {pages} {709} (\bibinfo {year} {1926})}\BibitemShut {NoStop}%
\bibitem [{\citenamefont {Batchelor}(1950)}]{batchelor_1950_similarity}%
  \BibitemOpen
  \bibfield  {author} {\bibinfo {author} {\bibfnamefont {G.~K.}\ \bibnamefont {Batchelor}},\ }\bibfield  {title} {\bibinfo {title} {The application of the similarity theory of turbulence to atmospheric diffusion},\ }\href {https://doi.org/10.1002/qj.49707632804} {\bibfield  {journal} {\bibinfo  {journal} {Quat. J. R. Meteor. Soc.}\ }\textbf {\bibinfo {volume} {76}},\ \bibinfo {pages} {133} (\bibinfo {year} {1950})}\BibitemShut {NoStop}%
\bibitem [{\citenamefont {Batchelor}(1952)}]{batchelor_1952_relative}%
  \BibitemOpen
  \bibfield  {author} {\bibinfo {author} {\bibfnamefont {G.~K.}\ \bibnamefont {Batchelor}},\ }\bibfield  {title} {\bibinfo {title} {Diffusion in a field of homogeneous turbulence. {II}. the relative motion of particles},\ }\href {https://doi.org/10.1017/S0305004100027687} {\bibfield  {journal} {\bibinfo  {journal} {Proc. Cambr. Phil. Soc.}\ }\textbf {\bibinfo {volume} {48}},\ \bibinfo {pages} {345} (\bibinfo {year} {1952})}\BibitemShut {NoStop}%
\bibitem [{\citenamefont {Konstandin}\ \emph {et~al.}(2012)\citenamefont {Konstandin}, \citenamefont {Federrath}, \citenamefont {Klessen},\ and\ \citenamefont {Schmidt}}]{Konstandin2012}%
  \BibitemOpen
  \bibfield  {author} {\bibinfo {author} {\bibfnamefont {L.}~\bibnamefont {Konstandin}}, \bibinfo {author} {\bibfnamefont {C.}~\bibnamefont {Federrath}}, \bibinfo {author} {\bibfnamefont {R.~S.}\ \bibnamefont {Klessen}},\ and\ \bibinfo {author} {\bibfnamefont {W.}~\bibnamefont {Schmidt}},\ }\bibfield  {title} {\bibinfo {title} {{Statistical properties of supersonic turbulence in the Lagrangian and Eulerian frameworks}},\ }\href {https://doi.org/10.1017/jfm.2011.503} {\bibfield  {journal} {\bibinfo  {journal} {J. Fluid Mech.}\ }\textbf {\bibinfo {volume} {692}},\ \bibinfo {pages} {183} (\bibinfo {year} {2012})}\BibitemShut {NoStop}%
\bibitem [{\citenamefont {Arn{\`e}odo}\ \emph {et~al.}(2008)\citenamefont {Arn{\`e}odo}, \citenamefont {Benzi}, \citenamefont {Berg}, \citenamefont {Biferale}, \citenamefont {Bodenschatz}, \citenamefont {Busse}, \citenamefont {Calzavarini}, \citenamefont {Castaing}, \citenamefont {Cencini}, \citenamefont {Chevillard}, \citenamefont {Fisher}, \citenamefont {Grauer}, \citenamefont {Homann}, \citenamefont {Lamb}, \citenamefont {Lanotte}, \citenamefont {L{\'e}v{\`e}que}, \citenamefont {L{\"u}thi}, \citenamefont {Mann}, \citenamefont {Mordant}, \citenamefont {M{\"u}ller}, \citenamefont {Ott}, \citenamefont {Ouellette}, \citenamefont {Pinton}, \citenamefont {Pope}, \citenamefont {Roux}, \citenamefont {Toschi}, \citenamefont {Xu},\ and\ \citenamefont {Yeung}}]{Arneodo2008}%
  \BibitemOpen
  \bibfield  {author} {\bibinfo {author} {\bibfnamefont {A.}~\bibnamefont {Arn{\`e}odo}}, \bibinfo {author} {\bibfnamefont {R.}~\bibnamefont {Benzi}}, \bibinfo {author} {\bibfnamefont {J.}~\bibnamefont {Berg}}, \bibinfo {author} {\bibfnamefont {L.}~\bibnamefont {Biferale}}, \bibinfo {author} {\bibfnamefont {E.}~\bibnamefont {Bodenschatz}}, \bibinfo {author} {\bibfnamefont {A.}~\bibnamefont {Busse}}, \bibinfo {author} {\bibfnamefont {E.}~\bibnamefont {Calzavarini}}, \bibinfo {author} {\bibfnamefont {B.}~\bibnamefont {Castaing}}, \bibinfo {author} {\bibfnamefont {M.}~\bibnamefont {Cencini}}, \bibinfo {author} {\bibfnamefont {L.}~\bibnamefont {Chevillard}}, \bibinfo {author} {\bibfnamefont {R.~T.}\ \bibnamefont {Fisher}}, \bibinfo {author} {\bibfnamefont {R.}~\bibnamefont {Grauer}}, \bibinfo {author} {\bibfnamefont {H.}~\bibnamefont {Homann}}, \bibinfo {author} {\bibfnamefont {D.}~\bibnamefont {Lamb}}, \bibinfo {author} {\bibfnamefont {A.~S.}\ \bibnamefont {Lanotte}}, \bibinfo {author} {\bibfnamefont
  {E.}~\bibnamefont {L{\'e}v{\`e}que}}, \bibinfo {author} {\bibfnamefont {B.}~\bibnamefont {L{\"u}thi}}, \bibinfo {author} {\bibfnamefont {J.}~\bibnamefont {Mann}}, \bibinfo {author} {\bibfnamefont {N.}~\bibnamefont {Mordant}}, \bibinfo {author} {\bibfnamefont {W.-C.}\ \bibnamefont {M{\"u}ller}}, \bibinfo {author} {\bibfnamefont {S.}~\bibnamefont {Ott}}, \bibinfo {author} {\bibfnamefont {N.~T.}\ \bibnamefont {Ouellette}}, \bibinfo {author} {\bibfnamefont {J.-F.}\ \bibnamefont {Pinton}}, \bibinfo {author} {\bibfnamefont {S.~B.}\ \bibnamefont {Pope}}, \bibinfo {author} {\bibfnamefont {S.~G.}\ \bibnamefont {Roux}}, \bibinfo {author} {\bibfnamefont {F.}~\bibnamefont {Toschi}}, \bibinfo {author} {\bibfnamefont {H.}~\bibnamefont {Xu}},\ and\ \bibinfo {author} {\bibfnamefont {P.~K.}\ \bibnamefont {Yeung}},\ }\bibfield  {title} {\bibinfo {title} {Universal intermittent properties of particle trajectories in highly turbulent flows},\ }\href {https://doi.org/10.1103/PhysRevLett.100.254504} {\bibfield  {journal}
  {\bibinfo  {journal} {Phys. Rev. Lett.}\ }\textbf {\bibinfo {volume} {100}},\ \bibinfo {pages} {254504} (\bibinfo {year} {2008})}\BibitemShut {NoStop}%
\bibitem [{\citenamefont {Ettel}\ \emph {et~al.}(2026)\citenamefont {Ettel}, \citenamefont {Samuel}, \citenamefont {Chertkov},\ and\ \citenamefont {Schumacher}}]{Ettel2026Lagrangian}%
  \BibitemOpen
  \bibfield  {author} {\bibinfo {author} {\bibfnamefont {M.}~\bibnamefont {Ettel}}, \bibinfo {author} {\bibfnamefont {R.~J.}\ \bibnamefont {Samuel}}, \bibinfo {author} {\bibfnamefont {M.}~\bibnamefont {Chertkov}},\ and\ \bibinfo {author} {\bibfnamefont {J.}~\bibnamefont {Schumacher}},\ }\bibfield  {title} {\bibinfo {title} {Lagrangian single-particle, multi-particle and topological analyses in turbulent {Rayleigh--B{\'e}nard} convection},\ }\bibfield  {journal} {\bibinfo  {journal} {arXiv preprint arXiv:2605.22326}\ }\href {https://doi.org/10.48550/arXiv.2605.22326} {10.48550/arXiv.2605.22326} (\bibinfo {year} {2026})\BibitemShut {NoStop}%
\bibitem [{\citenamefont {Schneide}\ \emph {et~al.}(2018)\citenamefont {Schneide}, \citenamefont {Pandey}, \citenamefont {Padberg-Gehle},\ and\ \citenamefont {Schumacher}}]{Schneide_PRF2018}%
  \BibitemOpen
  \bibfield  {author} {\bibinfo {author} {\bibfnamefont {C.}~\bibnamefont {Schneide}}, \bibinfo {author} {\bibfnamefont {A.}~\bibnamefont {Pandey}}, \bibinfo {author} {\bibfnamefont {K.}~\bibnamefont {Padberg-Gehle}},\ and\ \bibinfo {author} {\bibfnamefont {J.}~\bibnamefont {Schumacher}},\ }\bibfield  {title} {\bibinfo {title} {{Probing turbulent superstructures in {Rayleigh--B{'e}nard} convection by Lagrangian trajectory clusters}},\ }\href {https://doi.org/10.1103/PhysRevFluids.3.113501} {\bibfield  {journal} {\bibinfo  {journal} {Phys. Rev. Fluids}\ }\textbf {\bibinfo {volume} {3}},\ \bibinfo {pages} {113501} (\bibinfo {year} {2018})}\BibitemShut {NoStop}%
\bibitem [{\citenamefont {Vieweg}\ \emph {et~al.}(2021)\citenamefont {Vieweg}, \citenamefont {Schneide}, \citenamefont {Padberg-Gehle},\ and\ \citenamefont {Schumacher}}]{Vieweg_PRFLetter2021}%
  \BibitemOpen
  \bibfield  {author} {\bibinfo {author} {\bibfnamefont {P.~P.}\ \bibnamefont {Vieweg}}, \bibinfo {author} {\bibfnamefont {C.}~\bibnamefont {Schneide}}, \bibinfo {author} {\bibfnamefont {K.}~\bibnamefont {Padberg-Gehle}},\ and\ \bibinfo {author} {\bibfnamefont {J.}~\bibnamefont {Schumacher}},\ }\bibfield  {title} {\bibinfo {title} {Lagrangian heat transport in turbulent three-dimensional convection},\ }\href {https://doi.org/10.1103/PhysRevFluids.6.L041501} {\bibfield  {journal} {\bibinfo  {journal} {Phys. Rev. Fluids}\ }\textbf {\bibinfo {volume} {6}},\ \bibinfo {pages} {L041501} (\bibinfo {year} {2021})}\BibitemShut {NoStop}%
\bibitem [{\citenamefont {Schneide}\ and\ \citenamefont {Padberg-Gehle}(2017)}]{Schneide2017}%
  \BibitemOpen
  \bibfield  {author} {\bibinfo {author} {\bibfnamefont {C.}~\bibnamefont {Schneide}}\ and\ \bibinfo {author} {\bibfnamefont {K.}~\bibnamefont {Padberg-Gehle}},\ }\bibfield  {title} {\bibinfo {title} {{Network-based study of Lagrangian transport and mixing}},\ }\href@noop {} {\bibfield  {journal} {\bibinfo  {journal} {Nonlin. Processes Geophys.}\ }\textbf {\bibinfo {volume} {24}},\ \bibinfo {pages} {661} (\bibinfo {year} {2017})}\BibitemShut {NoStop}%
\bibitem [{\citenamefont {Fiedler}(1973)}]{Fiedler1973}%
  \BibitemOpen
  \bibfield  {author} {\bibinfo {author} {\bibfnamefont {M.}~\bibnamefont {Fiedler}},\ }\bibfield  {title} {\bibinfo {title} {Algebraic connectivity of graphs},\ }\href {https://doi.org/10.21136/CMJ.1973.101168} {\bibfield  {journal} {\bibinfo  {journal} {Czech. Math. J.}\ }\textbf {\bibinfo {volume} {23}},\ \bibinfo {pages} {298} (\bibinfo {year} {1973})}\BibitemShut {NoStop}%
\bibitem [{\citenamefont {von Luxburg}(2007)}]{Luxburg2007}%
  \BibitemOpen
  \bibfield  {author} {\bibinfo {author} {\bibfnamefont {U.}~\bibnamefont {von Luxburg}},\ }\bibfield  {title} {\bibinfo {title} {A tutorial on spectral clustering},\ }\href {https://doi.org/10.1007/s11222-007-9033-z} {\bibfield  {journal} {\bibinfo  {journal} {Stat. Comput.}\ }\textbf {\bibinfo {volume} {17}},\ \bibinfo {pages} {395} (\bibinfo {year} {2007})}\BibitemShut {NoStop}%
\bibitem [{\citenamefont {K{\"a}ufer}\ and\ \citenamefont {Cierpka}(2024)}]{Kaeufer2024Volumetric}%
  \BibitemOpen
  \bibfield  {author} {\bibinfo {author} {\bibfnamefont {T.}~\bibnamefont {K{\"a}ufer}}\ and\ \bibinfo {author} {\bibfnamefont {C.}~\bibnamefont {Cierpka}},\ }\bibfield  {title} {\bibinfo {title} {{Volumetric Lagrangian temperature and velocity measurements with thermochromic liquid crystals}},\ }\href {https://doi.org/10.1088/1361-6501/ad16d1} {\bibfield  {journal} {\bibinfo  {journal} {Meas. Sci. Technol.}\ }\textbf {\bibinfo {volume} {35}},\ \bibinfo {pages} {035301} (\bibinfo {year} {2024})}\BibitemShut {NoStop}%
\bibitem [{\citenamefont {K{\"a}hler}\ \emph {et~al.}(2012)\citenamefont {K{\"a}hler}, \citenamefont {Scharnowski},\ and\ \citenamefont {Cierpka}}]{Kahler2012Resolution}%
  \BibitemOpen
  \bibfield  {author} {\bibinfo {author} {\bibfnamefont {C.~J.}\ \bibnamefont {K{\"a}hler}}, \bibinfo {author} {\bibfnamefont {S.}~\bibnamefont {Scharnowski}},\ and\ \bibinfo {author} {\bibfnamefont {C.}~\bibnamefont {Cierpka}},\ }\bibfield  {title} {\bibinfo {title} {On the resolution limit of digital particle image velocimetry},\ }\href {https://doi.org/10.1007/s00348-012-1280-x} {\bibfield  {journal} {\bibinfo  {journal} {Exp. Fluids}\ }\textbf {\bibinfo {volume} {52}},\ \bibinfo {pages} {1629} (\bibinfo {year} {2012})}\BibitemShut {NoStop}%
\end{thebibliography}%

\begin{appendix}
\section{\label{sec:calibration}Calibration setup}
\begin{figure}
    \centering
    \includegraphics[width=0.4\linewidth]{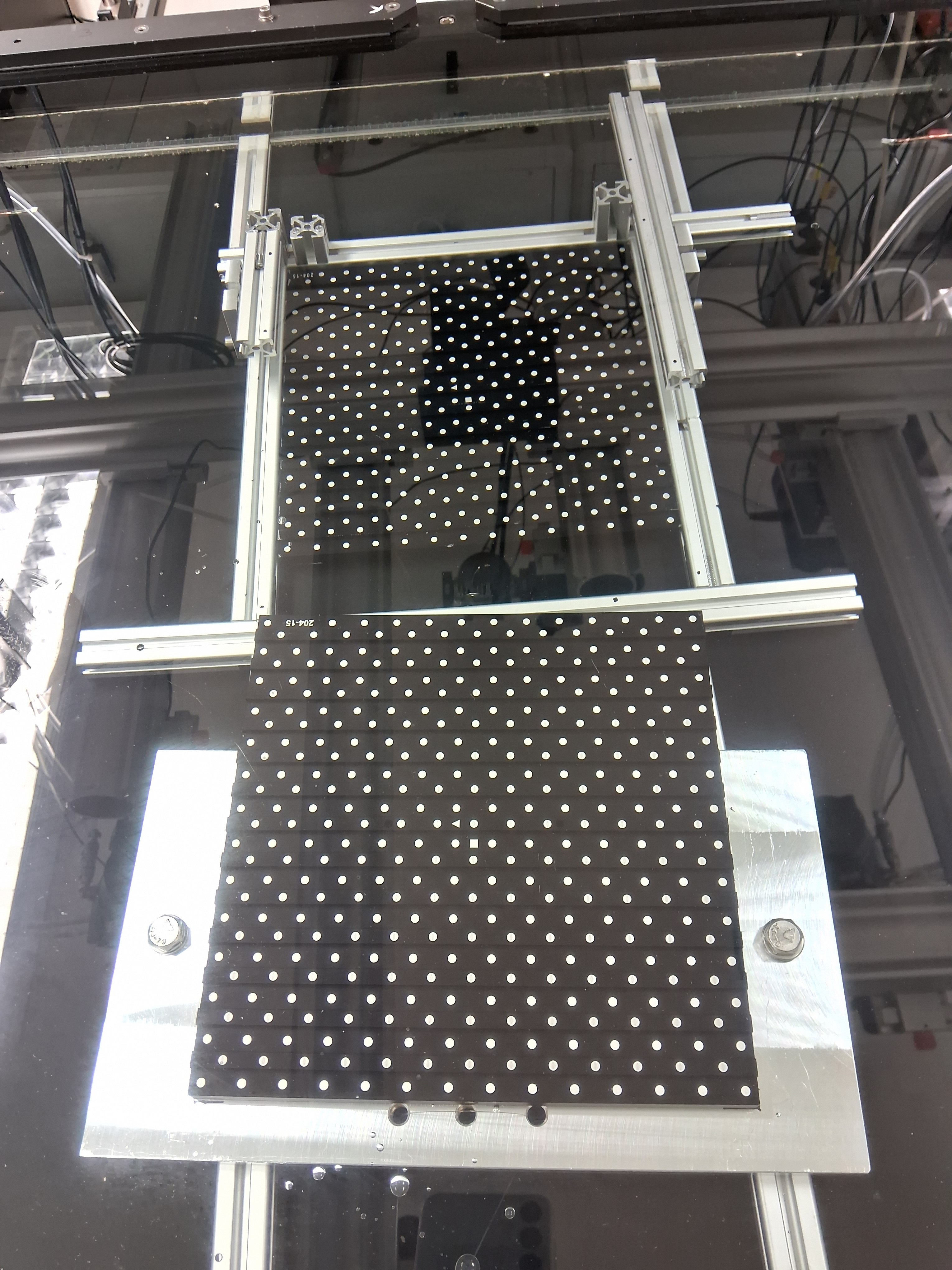}
    \caption{Two-view calibration setup components inside the convection cell captured from top using mobile camera.}
    \label{fig:calibration_setup}
\end{figure}

A calibration setup was built in-house using two 3D calibration plates (No. 204-15, LaVision GmbH), as shown in Fig.~\ref{fig:calibration_setup}. The setup was mounted on two precision rails and secured between the sides of the cell using nuts to prevent movement during the calibration procedure. Both plates were aligned parallel to the heating plate and to each other by means of a mounting arrangement incorporating precision rails. The top surfaces of the bottom and top calibration plates were located 17.4~mm and 54.2~mm from the heating plate, respectively. Thus, the bottom and top calibration plates were positioned approximately at the middle of the lower and upper halves of the convection cell, respectively. When the assembly was outside the cell, the distance between the two plates was measured at three different dot locations using a level sensor and was found to be 36.8$\pm$0.02~mm. Although Fig.~\ref{fig:calibration_setup} shows the top calibration plate resting on the side supports, at the start of the calibration process it was placed on top of the bottom calibration plate. The two plates are mechanically linked through precision rails to ensure vertical alignment of their edges, such that the dot patterns were positioned directly above one another. After recording the dot pattern on the top plate, it was removed from the field of view by means of a nylon thread attached to the top plate assembly and passed through a connecting tube on the side wall of the convection cell, enabling withdrawal from outside the cell. After the top plate was removed from the field of view, the bottom plate became visible in the cameras' field of view, and its dot pattern was subsequently recorded.
\begin{figure*}
\centering
  {\includegraphics[width=0.48\linewidth,trim=0 0 0 0,clip]{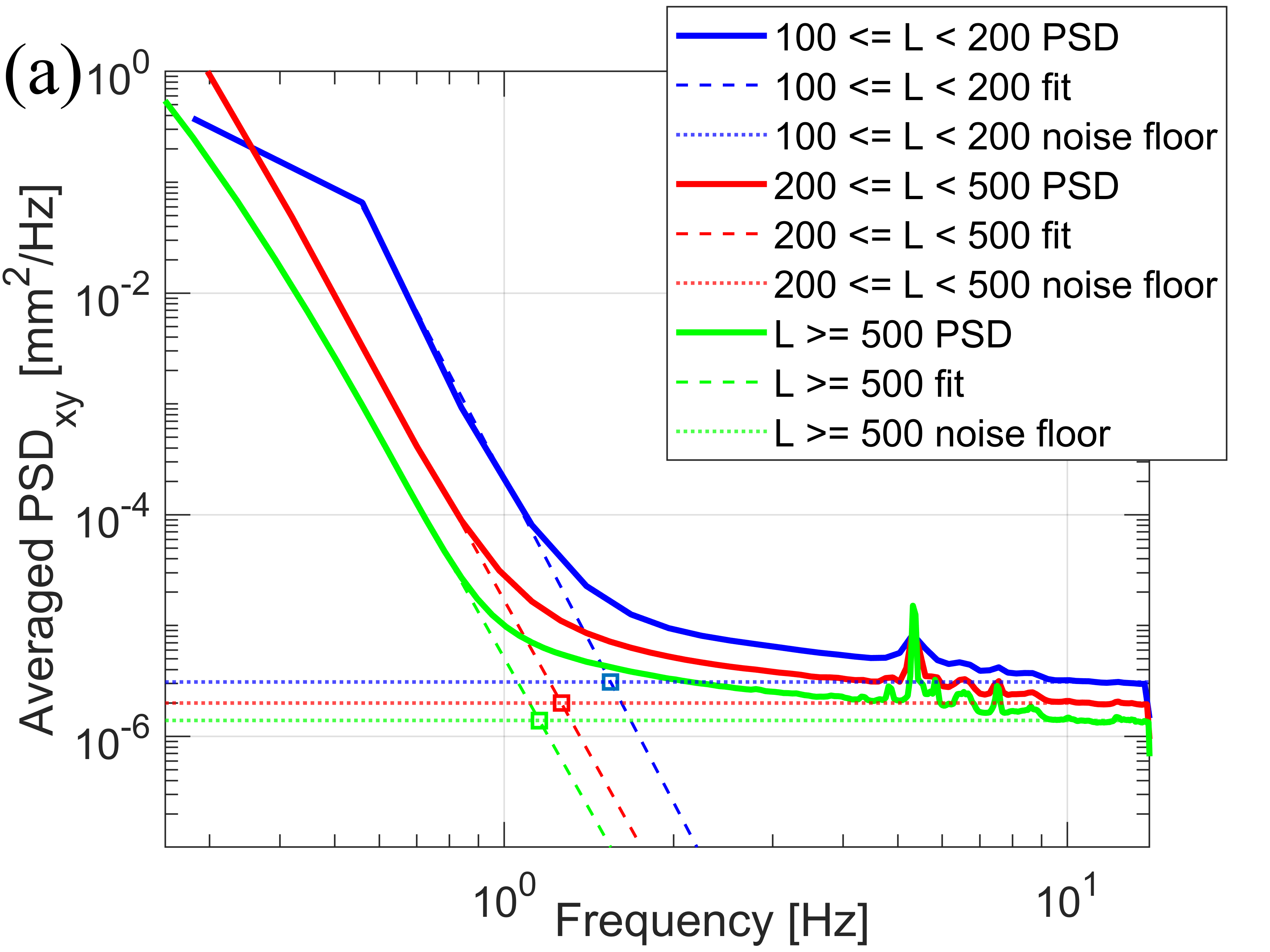}}
     {\includegraphics[width=0.48\linewidth,trim=0 0 0 0,clip]{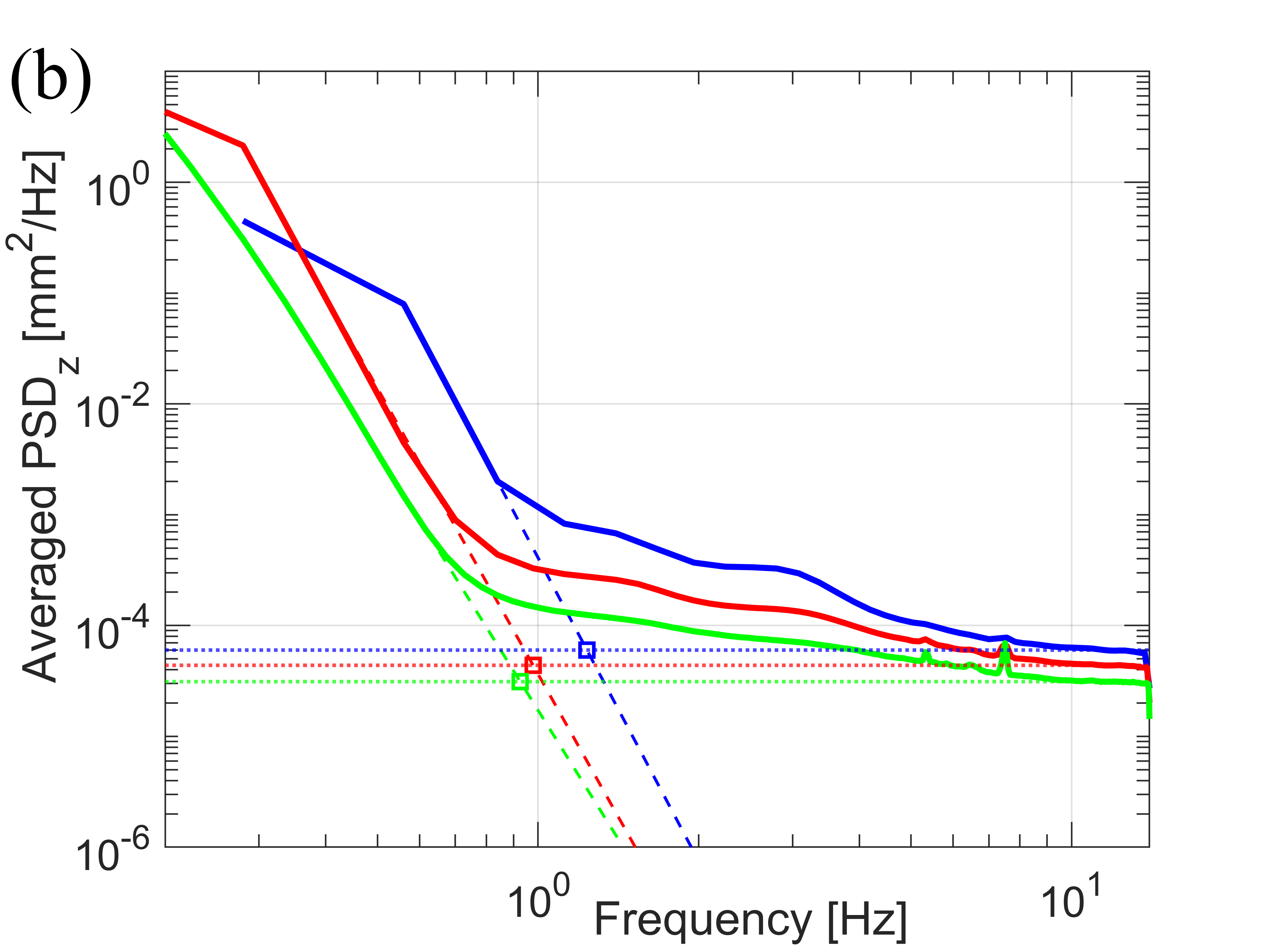}}\\
  \caption{Top panel: Ensemble averaged power spectral density of particle positions obtained from the raw STB trajectories of the lateral position components together, ($\rm{PSD}_{xy}$) (a) and of the vertical position component, ($\rm{PSD}_z$) (b), for all particle-tracking datasets acquired at 28 Hz. Each plot contains three PSD curves corresponding to three groups defined by particle track length ($L$). The plots also show the low-frequency PSD fit lines, the estimated noise floors, and the marked intersections between the fit lines and the corresponding noise floors, defined as cross-over points and the corresponding frequencies as cut-off frequencies ($f_c)$. The estimated cut-off frequencies for groups 1, 2, and 3 are given by \(\left(f_c^{xy}, f_c^{z}\right) = (1.5456, 1.2347), (1.2636, 0.9787), (1.1551, 0.9261)\), respectively.
  }
\label{fig:PSD}
\end{figure*}

\section{\label{sec:psd}Selection of cut-off frequency}
We obtained raw particle positions from STB for large number of particle trajectories. These trajectories had different lengths: some contained on the order of 100 particle positions, whereas others contained as many as 3000. It is known that faster particles are tracked for shorter durations, while slower particles are tracked for much longer periods~\cite{Wang_HIT2026}. In our analysis, we grouped the tracks into three categories according to their lengths: (1) tracks with lengths ($L \leq 199$), (2) tracks with ($200 \leq L < 499$), and (3) tracks with ($L \geq 500$). Tracks in groups 1, 2, and 3 were trimmed to lengths of 100, 200, and 500, respectively, to compute the ensemble averaged PSD for each group. Furthermore, the tracks were also grouped according to the lateral and vertical directions. Figures~\ref{fig:PSD}(a,b) show the ensemble averaged power spectral density (PSD) curves computed from the particle-tracks in the lateral and vertical directions, respectively. Owing to measurement noise from various sources, all curves flatten beyond a certain frequency. To estimate the noise-floor level, we trim the edges of each spectrum and average the final 20$\%$ of the remaining spectral values. A horizontal line at the noise-floor level is drawn. In addition, the low-frequency, energy-containing region is fitted with a straight line, and the intersection of this fit with the horizontal noise-floor line is used to define the cut-off frequency $f_c$ for each curve individually. The noise-floor line, the fitted low-frequency line, and their intersection points are shown in Figs.~\ref{fig:PSD}(a) and (b). The resulting cut-off frequencies are then used to filter the trajectories within their respective groups. The intercepts of the noise-floor lines with the ordinate axis indicate that the lateral-position data have higher spatial resolution than the vertical-position data. The position measurement errors were 7.5 $\mu\text{m}$ and 26 $\mu\text{m}$ in the lateral and vertical directions, respectively. With a camera scale factor of 16.33 pixels per mm, these errors correspond to particle displacements of approximately 0.11 and 0.37 pixels in the lateral and vertical directions, respectively. These values are close to the minimum achievable resolution of PTV~\cite{Kahler2012Resolution} and are comparable to the maximum disparity values obtained through volume self-calibration. Furthermore, these error values are much smaller than $\eta_k$~(Table~\ref{tab:expt_para}), which corresponds to a particle displacement of 11 pixels. Moreover, shorter tracks exhibit higher cut-off frequencies, whereas the cut-off frequencies in the vertical direction are lower than those in the lateral direction. We fit the measured positions using cubic B-splines with a penalty imposed on the third derivative (jerk). Specifically, we minimise the cost function
\begin{equation}
\label{eq}
J[x(t)] =
\sum_{i=1}^{n}\left|x(t_i)-Q(t_i)\right|^2
+
\sum_{i=1}^{n-1}\left| \lambda.
x^{(3)}\left( t_{j+\frac{1}{2}}\right)
\right|^2,
\end{equation}
where $x(t_i)$ denotes the corrected spline position at time $t_i$, $Q(t_i)$ the measured position, and $x^{(3)}(t)$ the third derivative of the corrected trajectory. The parameter $\lambda$ is defined as $
\lambda=1/(2\pi f_c/f_s)^3,
$
where $f_s$ is the sampling frequency and $f_c$ the cross-over frequency corresponds to signal to noise ratio of one.

Using the filtered tracks, we found that the shorter trajectories exhibit substantially higher vertical root-mean-square accelerations, which may explain the observed shift in the PSD curves.
These trajectories are likely shorter because particles with larger accelerations are more challenging for the tracking algorithm to follow over long durations. 
\end{appendix}
\end{document}